\documentclass[twocolumn]{emulateapj}
\usepackage{graphicx}
\usepackage[dvips]{color}
\usepackage{amsmath,amsthm,amssymb}
\usepackage{mathtools}

\newcommand{\be}{\begin{equation}}
\newcommand{\ee}{\end{equation}}
\newcommand{\ba}{\begin{eqnarray}}
\newcommand{\ea}{\end{eqnarray}}

\def\lsim{\raise0.3ex\hbox{$\;<$\kern-0.75em\raise-1.1ex\hbox{$\sim\;$}}}
\def\gsim{\raise0.3ex\hbox{$\;>$\kern-0.75em\raise-1.1ex\hbox{$\sim\;$}}}

\def\theta{\vartheta}

\shortauthors{Giacinti \& Kirk}
\shorttitle{Electron Acceleration in the Crab Nebula}

\begin{document}

\author{Gwenael~Giacinti and John~G.~Kirk}
\affil{Max-Planck-Institut f\"ur Kernphysik, Postfach 103980, 69029 Heidelberg, Germany}

\title{Acceleration of X-ray Emitting Electrons in the Crab Nebula}

\begin{abstract}
  We study particle acceleration at the termination shock of a striped
  pulsar wind by integrating trajectories in a prescribed model of the
  magnetic field and flow pattern.  Drift motion on the shock surface
  maintains either electrons or positrons on \lq\lq Speiser\rq\rq\
  orbits in a ring-shaped region close to the equatorial plane of the
  pulsar, enabling them to be accelerated to very high energy by the
  first-order Fermi mechanism.  A power-law spectrum results: $dN_{\rm
    e}/d\gamma\propto\gamma^{\alpha_{\rm e}}$, where $\alpha_{\rm e}$
  lies in the range $-1.8$~to~$-2.4$ and depends on the downstream
  turbulence level. For sufficiently strong turbulence, we find
  $\alpha_{\rm e}\simeq -2.2$, and both the photon index and the flux
  of $1$--$100$\,keV X-rays from the Crab Nebula, as measured by
  NuSTAR, can be reproduced. The particle spectrum hardens to
  $\alpha_{\rm e} \simeq -1.8$ at lower turbulence levels, which may
  explain the hard photon index observed by the Chandra X-ray
  Observatory in the central regions of the Nebula.
  
\end{abstract}


\maketitle

\section{Introduction}
\label{Introduction}

The photon index, $\Gamma=2.1$, of the Crab Nebula in $1$--$100\,$keV
X-rays \citep{NuSTAR2015} is very close to that predicted for
electrons accelerated by the first-order Fermi process at a
relativistic shock front
\citep{bednarzostrowski98,kirketal00,Achterberg2001}.  Is this just a
coincidence? On the one hand, this mechanism is known to be inhibited
at perpendicular shocks \citep{begelmankirk90,sironispitkovsky09,summerlinbaring12},
such as that separating the pulsar wind from the Crab Nebula. The
reason is that the magnetic field sweeps particles away from the shock
in the downstream region, thereby preventing the multiple, stochastic 
shock crossings that characterize the Fermi process.  On the other
hand, the toroidal magnetic field transported through the shock into
the Nebula is expected to change sign across the rotational equatorial
plane of the pulsar \citep[for reviews, see][]{amato14,porthetal17}, giving
rise to a broad current sheet, in which the Fermi process might still
operate. To answer the question posed above 
and determine the relevance of this process, we study particle
acceleration in the equatorial sheet using a detailed model of the
magnetic field there.  We find that stochastic crossings and
recrossings of the shock front are indeed responsible for
acceleration, and that shock-induced drifts play a crucial role in
focusing leptons of one sign of charge into the acceleration
zone. Our main result is that both the photon index and the flux of
X-rays can be reproduced by the combination of Fermi acceleration and
drifts, if one assumes a turbulent amplitude $\delta B_{\rm
  d}>200\,\mu$G and an average toroidal field at higher latitudes of
$B=1\,$mG.

Recent, state-of-the-art, phenomenological modeling of the morphology
of the Crab Nebula places significant constraints on the possible
sites of particle acceleration. In particular, the X-ray to soft
gamma-ray emission appears to originate from a torus-shaped region
lying in the rotational equator of the Crab Pulsar, and located at a
radius where the ram pressure of the pulsar wind roughly equals that
in the Nebula \citep{porthetal14,olmietal15}. Furthermore, these
models give insight into the global structure of the magnetic field 
and the degree to which it is turbulent, making it possible to
construct diffusion coefficients for energetic particles propagating
in the outer Nebula \citep{Porth2016}. However, close to the
relativistic termination shock (TS) that forms the inner edge of the
Nebula, the energetic particle distribution is necessarily anisotropic
\citep{kirkschneider87}, so that diffusion coefficients cannot be used
to model the transport process. Instead, we build a simplified, 
explicit model of the magnetic field and flow pattern in the
equatorial region of the TS, based on the results of
MHD simulations, and follow the trajectories of particles injected
at the shock as they cross and recross it. Finally, we compute the
radiation they emit when cooling in the Nebula, after leaving the shock.

The magnetic field model, injection prescription, and method of
computing the radiation are described in \S\ref{Model}, and the
results found by analyzing particle trajectories in \S\ref{Results}. A
discussion of the application to the Crab Nebula is presented in
\S\ref{Discussion}.

\section{Description of the model}
\label{Model}

\subsection{Regular magnetic field}
Magnetohydrodynamic models of the Crab Nebula suggest that it is
powered by a radially propagating pulsar wind, whose luminosity per
unit solid angle is concentrated towards the rotational equator. The
particle component, which we assume to be electrons and positrons,
carries only a small fraction of the power close to launch, most of it
being in the form of Poynting flux. However, the wind is thought
to be striped \citep{coroniti90,michel94}, i.e., the magnetic field has a
component that oscillates at the rotation frequency of the pulsar, as
well as a phase-averaged or DC component.  MHD models assume complete
dissipation of the oscillating component before the plasma enters the
Nebula downstream of the TS 
\citep{delzannaetal18}. Whether this occurs somewhere in
the wind, or at the shock itself, has no influence on the downstream
parameters, provided it proceeds without significant radiation losses
\citep{lyubarsky03}.  
The remaining, phase-independent magnetic field is
carried into the Nebula, and reverses its sign across the rotational
equator. Thus, an equatorial current sheet is formed, whose thickness depends on
the latitude distribution of the oscillations, which, in turn, is
determined by the inclination angle between the magnetic and rotation
axes of the pulsar. The TS itself is oblate: in the
equatorial region, it is approximately spherical with radius roughly
$4\times10^{17}\,$cm, but it moves close in to the pulsar in the polar
regions, where the power of the wind is low.  Outside the current
sheet, but still in the equatorial region, the ordered field is
roughly $1\,$mG on the downstream side of the shock.

\begin{figure*}
  \includegraphics[width=0.99\textwidth]{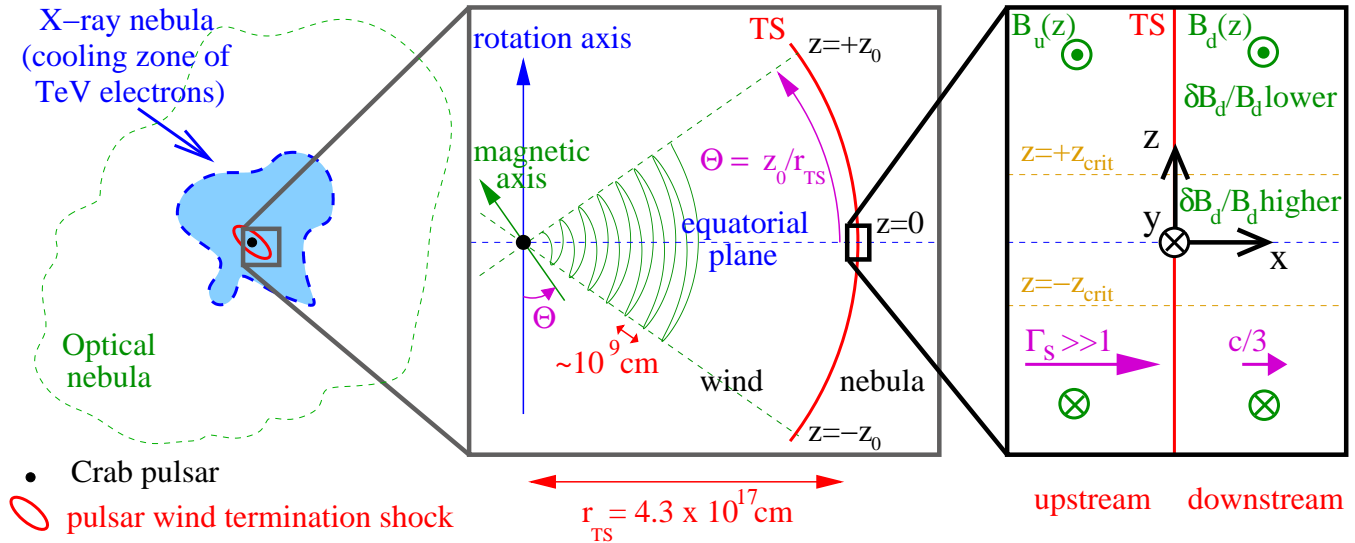}
  \caption{ Sketch of the Crab Nebula and the location of the
    equatorial acceleration region.  In the left panel, the outlines of
    the optical and X-ray nebulae are shown as they appear on the sky
    \citep[e.g., Fig.~3 in][]{hester08}, together with an estimate of
    the position of the termination shock of the wind, drawn roughly
    to scale. The center panel shows the equatorial region of the
    shock oriented such that the rotation axis of the pulsar is in the
    vertical direction and sketched at a phase at which the magnetic
    axis is in the plane of the figure. Magnetic field oscillations
    (stripes) of wavelength $c$ times the pulsar rotation period are
    present upstream of the TS in a sector around the equator, and are
    depicted on a greatly expanded length scale. Both upstream and
    downstream, the phase-averaged, toroidal field reverses sign
    across the equator. In the right-hand panel, the region
    $\left|z\right|\lesssim z_{\rm crit}$ (defined in
    Eq.~(\ref{z_crit})) of the equatorial current sheet is shown, in
    which particles injected at the shock can be accelerated by the
    first-order Fermi process. Typically, this region makes up a few
    percent of the shock area (see Table~\ref{Table_Whole_TS}), i.e.,
    $r_{\rm TS}\gtrsim z_0\gtrsim 10\times z_{\rm crit}$.}
\label{Sketch}
\end{figure*}

Here we adopt a planar model of the flow in the equatorial region,
since the gyroradius of TeV--PeV electrons in a mG magnetic field is
$\simeq 3 \times 10^{12 - 15}$\,cm, much smaller than the radius of
the shock. In cartesian coordinates, the shock is located in the $x=0$
plane, and the equatorial plane is $z=0$. The location of this region 
with respect to the observed optical and X-ray nebulae is shown in 
Fig.~\ref{Sketch}.
In the downstream region,
$x>0$, the plasma is assumed to flow everywhere along $+{\bf \hat{x}}$
at $c/3$, as expected behind a strong, weakly magnetized, relativistic
shock, and the current sheet is located at $-z_0<z<z_0$.  The
downstream magnetic field, ${\bf B_{\rm d}}(z)$, measured in the rest
frame of the downstream fluid (DRF), is linearly interpolated between
the values on the northern and southern edges of the current sheet:
\begin{equation}
 {\bf B_{\rm d}}(z) = \left\{ \begin{array}{ll}
	      	    -B_{\rm d,0} {\bf \hat{y}} & {\rm ~if~} z > z_{0}\\
	      	    -B_{\rm d,0} (z/z_{0}) {\bf \hat{y}} & {\rm ~if~} |z| \leq z_{0}\\
	      	    +B_{\rm d,0} {\bf \hat{y}} & {\rm ~if~} z < -z_{0} \;.
                    \end{array} \right.
\label{B_d}
\end{equation}

To find the corresponding field in the upstream region, we assume that the
TS is a thin structure in which all incoming
oscillations at the pulsar rotation frequency are dissipated.
Applying Faraday's law, together with a time-average over the pulsar
period, one finds that the electric and magnetic fields upstream, $
{\bf E'_{\rm u}}(z)$ and $ {\bf B'_{\rm u}}(z)$, as measured in the
shock rest frame (the \lq\lq SRF\rq\rq), are:
\begin{equation}
 {\bf E'_{\rm u}}(z) =\frac{1}{2\sqrt{2}}\times \left\{ \begin{array}{ll}
	      	    +B_{\rm d,0} {\bf \hat{z}} & {\rm ~if~} z > z_{0}\\
	      	    +B_{\rm d,0} (z/z_{0}) {\bf \hat{z}} & {\rm ~if~} |z| \leq z_{0}\\
	      	    -B_{\rm d,0} {\bf \hat{z}} & {\rm ~if~} z < -z_{0}
                    \end{array} \right. 
\label{Eprime_u}
\end{equation}
\begin{equation}
 {\bf B'_{\rm u}}(z) =\frac{1}{2\sqrt{2}\beta_{\rm s}}\times \left\{ \begin{array}{ll}
	      	    -B_{\rm d,0} {\bf \hat{y}} & {\rm ~if~} z > z_{0}\\
	      	    -B_{\rm d,0} (z/z_{0}) {\bf \hat{y}} & {\rm ~if~} |z| \leq z_{0}\\
	      	    +B_{\rm d,0} {\bf \hat{y}} & {\rm ~if~} z < -z_{0}
                    \end{array} \right.
\label{Bprime_u}
\end{equation}
where $\beta_{\rm s}{\bf \hat{x}}$ is the 3-velocity of the upstream
plasma in the SRF. Thus, for highly relativistic inflow, $\beta_{\rm
  s}\approx1$, the fields seen in the shock frame are, to a good
approximation, equal to those of a vacuum electromagnetic wave.  It
follows that the particle trajectories are insensitive to the Lorentz
factor $\Gamma_{\rm s}=1/\sqrt{1-\beta_{\rm s}^2}$ of the upstream
plasma. The oscillating component is not constrained by this analysis,
but this is not important in the present context, since the gyroradius
of particles injected into the acceleration process in the equatorial
zone substantially exceeds the wavelength of the oscillations, which,
therefore, provide only a small perturbation of the orbit computed in
the phase-averaged field.

In the absence of turbulence, particles far from the equator undergo
systematic drifts in either the positive or negative 
${\bf x}$-direction, superimposed upon the plasma bulk motion.  Provided the
drift motion in the plasma rest frame is slower than the plasma speed 
in the SRF, which is always true in
the cases we consider, all particles move in the direction of the
flow, i.e., towards the shock in the upstream, and away from it in the downstream
region.  However, a crucial, novel aspect is introduced by the
reversal of the average field. Speiser orbits \citep{speiser65}, which
cross the plane $z=0$, do not drift, but can propagate at arbitrary
speed (consistent with their energy) in the $\pm{\bf x}$
directions. Thus, there exists a population of particles that is
effectively disconnected from the local plasma speed, which
facilitates repeated shock crossings. 
As a rough guide, a particle of energy $E_{\rm inj,d}$ injected into the field defined in Eq.~(\ref{B_d})
at height $z$ above the equatorial plane follows a Speiser orbit if $\left|z\right|<z_{\rm crit}$, where
\begin{equation}
  z_{\rm crit} = \sqrt{\frac{z_{0}E_{\rm inj,d}}{eB_{\rm d,0}}} 
\simeq 5.8 \times 10^{14}\,{\rm cm}\, 
\sqrt{\frac{z_{0,17}E_{\rm inj,d,12}}{B_{\rm d,0,-3}}}\;,
\label{z_crit}
\end{equation}
where $z_{0,17}=z_{0}/(10^{17}$\,cm), $B_{\rm d,0,-3}=B_{\rm
  d,0}/(1$\,mG), and $E_{\rm inj,d,12}=E_{\rm inj,d}/(1$\,TeV).

\subsection{Turbulent magnetic field}
\label{NumSim}

Onto the large-scale magnetic field ${\bf B_{\rm d}}(z)$, we 
superimpose a three-dimensional,
homogeneous turbulent field, $\delta{\bf B_{\rm d}}(x,y,z)$ 
(also defined in
the DRF). This field satisfies $\langle \delta{\bf B_{\rm d}}
\rangle = {\bf 0}$, where $\left<\dots\right>$ denotes a spatial average,
 and its root-mean-square strength, $\delta
B_{\rm d} \equiv (\langle \delta{\bf B_{\rm d}}^{2} \rangle
)^{1/2} > 0$, is independent of position.
This implies that the level of turbulence, defined as
$\delta B_{\rm d}/B_{\rm d}$ with $B_{\rm d}=|{\bf B_{\rm
    d}}|$, is larger at small $|z|$, in line with results
from MHD simulations of pulsar wind nebulae. See, for instance, the
upper right panel in Fig.~4 of~\cite{Porth2016}, where the largest
levels of turbulence in the TS downstream are obtained around the
equatorial plane. 

We generate $\delta{\bf B_{\rm d}}$ on
3D grids with $\mathcal{N}=256$ vertices per side ($256^{3}$ grid
points in total), following the method presented and tested
in~\cite{Giacinti2012}. The grids repeat periodically in
space, and the three cartesian components of $\delta{\bf B_{\rm d}}$
are calculated at any point in space using an eight-point linear
interpolation of their values on the eight nearest vertices of the
grid. We generate isotropic Bohm turbulence with power-spectrum
$\mathcal{P}(k) \propto k^{-1}$, for wave vectors in the range
$2\pi/L_{\max} \leq k \leq 2\pi/L_{\min}$, where $L_{\max}$ is the
lateral size of the grid and $L_{\min}$ is twice the spacing
between grid points. The dynamical range of the turbulence is,
therefore, $L_{\max}/L_{\min}=\mathcal{N}/2=128$. We choose the spacing
between grid points to be slightly smaller than half of the gyroradius 
in the strongest magnetic field in the injection zone 
of an electron with energy $E_{\rm inj,d}$ in the DRF. 
Taking smaller values does not
noticeably affect the results. The value of $L_{\max}$ determines the
high-energy cutoff in our simulated electron spectra,
because particles with gyroradii larger than $L_{\max}$ experience
little scattering and, therefore, no longer gain energy via the
first-order Fermi mechanism. We have also tested other
power-spectra, such as Kolmogorov ($\mathcal{P}(k) \propto k^{-5/3}$),
and found no significant difference. 

The Fermi process depends on the competing effects of 
advection and diffusion due to turbulence. Therefore, since 
particles injected at $\left|z\right| < z_{\rm crit}$ 
follow trajectories resembling Speiser orbits, whereas those
injected at $\left|z\right| > z_{\rm crit}$ are predominantly
advected with the plasma, differences 
can be expected according to whether the level of turbulence at 
$z_{\rm crit}$ is smaller or larger than unity. 
We denote the dimensionless parameter
characterizing these different acceleration regimes by
\begin{eqnarray}
\eta_{\rm crit} &\equiv& \delta B_{\rm d}/B_{\rm d} (z_{\rm crit})\,,
\label{etacritdef}
\end{eqnarray}
and investigate a range of values covering small and large $\eta_{\rm crit}$, 
whilst keeping the magnetic field
at $z\sim z_{0}$ predominantly toroidal, as indicated by
simulations \citep{Porth2016}. 

The idealized, plane-parallel case with only a phase-averaged field in
the upstream region introduces an unphysical feature into the particle
kinematics: it permits particles moving very close to the equator to
propagate unhindered to an arbitrarily large distance upstream of the
shock. In reality, both the spherical geometry and irregularities in
the oscillating and the phase-averaged fields prevent this
behavior. In our simulations, we take account of this by adding to
the upstream, phase-averaged component a small turbulent field that is
purely magnetic as seen in the upstream rest frame (URF), 
in analogy with that added to the
downstream field, but physically disconnected from it. This
turbulent component maintains the conservation of particle energy
measured in the URF, making it convenient to integrate the
trajectories in this reference frame. 
To ensure that particles moving almost along
$-{\bf \hat{x}}$ experience resonant scattering, we also stretch the
grid in the upstream by a factor $\Gamma_{\rm s}$ along $x$. We have
performed tests to ensure that the properties of this turbulent
upstream field do not affect our results.

\subsection{Injection}
In an isotropic wind, the energy carried per particle in units of
$m_{\rm e}c^{2}$, after dissipation of the entire Poynting flux, is
\begin{equation}
  \mu = \frac{L_{\rm s.d.}}{\dot{N}_{\pm}m_{\rm e}c^{2}}\;,
\label{Mu}
\end{equation}
where $L_{\rm s.d.}$ is the spin-down power of the neutron star, and
$\dot{N}_{\pm}$ the rate at which the particles are transported into
the nebula by the wind. In the absence of a phase-averaged field,
i.e., precisely on the equator, the results of \cite{amanokirk13},
\cite{giacchekirk17}, and \cite{Kirk2017}, indicate that particles are
effectively thermalized in a thin structure, termed an \lq\lq
electromagnetically modified shock front\rq\rq. The majority of the particles 
are transmitted through this structure into the downstream region with
energy in the DRF $E_{\rm inj,d}=\gamma_{\rm inj,d}m_{\rm
  e}c^{2}\approx \mu m_{\rm e}c^{2}$, and a small fraction is
reflected into the upstream region. To date, computations of the
shock structure with a {\em non}-vanishing phase-averaged field
\citep{sironispitkovsky11} are available only for a uniform field
and high plasma density --- a regime which is unlikely to be
relevant in the case of the Crab \cite[see the discussion in][]{amanokirk13}. 
The physics of the high-density, uniform field
case also differs significantly from that considered here,
because (i) the wavelength of the oscillations is much larger than
the relativistic Larmor radius of the upstream particles, (ii) the
shock does not undergo electromagnetic modification, and (iii)
particles cannot be reflected, because of the absence of Speiser
trajectories. Nevertheless, the PIC simulations cited above are in
good agreement with the simple estimate that the injection energy
equals the energy carried per particle after dissipation of the
oscillating component of the magnetic field.  Particles that undergo
acceleration are injected relatively close to the equator, with
$\left|z\right|/z_{0}\lesssim 0.1$ (see Table~\ref{Table_Whole_TS}),
where the energy density in the phase-averaged field ${\bf B}_{\rm d}^2/8\pi$ 
is less than roughly 1\,\% of the total energy
density. Therefore, independent of the precise position, we assume
particles are injected into the downstream plasma with the same
value of $E_{\rm inj,d}$ as at the equator. In addition, we assume
injected particles have an initial momentum directed along the shock
normal. On the one hand, these assumptions slightly overestimate the
injection energy at finite $\left|z\right|$, but, on the other, they
underestimate it by neglecting the reflected particles, and also
underestimate the initial return probability by assuming the angular
distribution of the injected particles to be a collimated beam.
The average value of $\mu$ over the entire lifetime of the Crab Nebula
and over all directions of the wind is constrained to be $10^{4}
\lesssim \mu \lesssim 10^{6}$ \citep{Olmi2016,porthetal17}; in our
simulations, we choose $E_{\rm inj,d}=1$\,TeV, corresponding to
$\gamma_{\rm inj,d} \approx 2 \times 10^{6}$.

\subsection{Simulated trajectories}

We integrate the particle trajectories in the test-particle limit, by
solving the Lorentz force equation in the upstream and downstream rest
frames where the electric fields vanish. Each time a particle crosses
the shock, a Lorentz transformation of the momentum components is
performed from the old rest frame to the new rest frame. Although it
does not affect the final result, this procedure requires a specific
choice of upstream Lorentz factor, for which we choose $\Gamma_{\rm
  s}=100$.  In the DRF (URF), the shock is located at $x_{\rm d} =
-ct_{\rm d}/3$ ($x_{\rm u} = -\beta_{\rm s}ct_{\rm u}$). We note
that advection of particles with the fluid flow is automatically taken
into account by this procedure. We place an escape boundary in the
downstream at $x=+d$, as measured in the SRF, and terminate each
trajectory when it reaches $x=d$. We have verified that the results do
not depend on $d$, provided it is larger than the gyroradius of the
highest energy electrons present in the system. On the upstream side,
particles cannot escape to $x \rightarrow -\infty$, because the shock
always overtakes them. At each shock crossing, all relevant physical
quantities of the accelerated particles (energies in the DRF and SRF,
momenta coordinates, positions, times) are stored. These are used, for
example, to calculate the steady-state spectra of the accelerated
electrons and positrons at the shock front.

\subsection{Synchrotron emission from the nebula}
\label{SynchCalc}

The particles accelerated at the TS are ultimately advected into the
nebula, where they cool and emit synchrotron radiation; 
see the area shaded in blue in the left panel of Fig.~\ref{Sketch}. In a magnetic
field $B$, the synchrotron power emitted per unit frequency interval
by a single electron with pitch angle $\alpha$ and nonrelativistic   
(angular) gyrofrequency $\omega_{\rm g}=eB/m_{\rm e}c$ is
\begin{equation}
  \frac{{\rm d} L_{\rm 1p}^{\rm synch}}{{\rm d} \nu} = \sqrt{3} \alpha_{\rm f}
\hbar\omega_{\rm g}\sin\alpha {\rm F} \left( \nu/\nu_{\rm c} \right)\;,
\label{Synch_1p}
\end{equation}
where $\alpha_{\rm f}$ is the fine-structure constant, 
$\nu_{\rm c} = 3\gamma^{2} \omega_{\rm g}\sin\alpha /(4\pi)$ is the characteristic 
frequency, and the synchrotron function is
\begin{equation}
  {\rm F} (x) = x \int_{x}^{\infty} {\rm d}t K_{5/3}(t)\;,
\label{F}
\end{equation}
where $K_{5/3}$ is a modified Bessel function. In the following, we
neglect the dependence on pitch angle by setting $\sin \alpha = \sqrt{2/3}$
and approximate the synchrotron function by ${\rm F} (x) = 1.85 \, x^{1/3} \exp
(-x)$~\citep[see][]{Melrose1980}. The resulting total luminosity per
unit frequency interval is
\begin{equation}
  J(\nu) = \int {\rm d}\gamma N_{\rm c}(\gamma) \frac{{\rm d} L_{\rm 1p}^{\rm synch}}{{\rm d} \nu}\;,
\label{J_Nu}
\end{equation}
where $N_{\rm c}(\gamma) = {\rm d}N_{\rm c}/{\rm d}\gamma$ is the
differential number of cooled electrons in the interval ${\rm
  d}\gamma$ in the nebula, and we have implicitly assumed a
homogeneous magnetic field within the radiation zone. For a source at
a distance $D$ from Earth, the differential energy flux is
$F_{\nu}=J(\nu)/(4\pi D^{2})$. Synchrotron losses imply 
$\dot{\gamma}=-\beta\gamma^{2}$ with $\beta=\sigma_{\rm
  T}B^{2}/(6\pi m_{\rm e}c)$ and $\sigma_{\rm T}$ the Thomson cross
section, and one finds, in the steady state regime,
\begin{equation}
  N_{\rm c}(\gamma) = \frac{1}{\beta\gamma^{2}}\,\int_{\gamma}^{\infty} {\rm d}\gamma^{\prime} Q(\gamma^{\prime}) \;,
\label{N_c}
\end{equation}
where $Q(\gamma){\rm d}\gamma$ is the number of particles
accelerated at the TS and \lq\lq injected\rq\rq\/ into the nebula per
time unit with a Lorentz factor between $\gamma$ and $\gamma+{\rm d}\gamma$. 

We set 
\begin{equation}
Q(\gamma_{\rm d})=\left\{ \begin{array}{ll}
                  Q_{0}\gamma_{\rm d}^{\alpha_{\rm e}} & \textrm{for\ }
f E_{\rm inj,d}/m_{\rm e}c^{2}
\leq \gamma_{\rm d} \leq E_{\max}/m_{\rm
  e}c^{2}\\
0                                                 & \textrm{otherwise}
                 \end{array} \right.
\label{defineq0}
\end{equation}
and determine the spectral index, $\alpha_{\rm e}$, 
from the results described in \S~\ref{Results_Equatorial_Plane}. 
The parameter $f$ is chosen such that the simulated particle spectrum at the TS
is a power-law at $E_{\rm d} \geq f \times E_{\rm inj,d}$. Typically, we find $f=3$ to 7.
Particles of energy less than $f \times E_{\rm inj,d}$ are neglected 
in (\ref{defineq0}), but they influence only the low-frequency 
synchrotron spectrum, $\nu\lesssim\left(fE_{\rm inj,d}/mc^2\right)^2\omega_{\rm g}$.
We do not attempt to model the spectrum of the Nebula in this energy range, since it is 
less well-known, because of the uncertainty 
associated with the contribution of the pulsar and 
the difficulties involved in modeling absorption \citep{kirschetal05}. 
The limited dynamical range of the turbulence in our simulation
introduces an artificial upper limit to the power law distribution of
accelerated particles. In reality, however, this quantity, 
$E_{\rm max}$, is determined by radiative losses, even though these can be
neglected over most of the acceleration range.
Setting the loss-time, $\tau_{\rm sync}=6\pi m_{\rm
  e}^{2}c^{3}/(\sigma_{\rm T}B^{2}E)$, equal to the time to complete
one half of a gyration, $\tau_{1/2}=\pi E/(eBc)$, at
$E=E_{\rm max}$, leads to
\begin{equation}
E_{\max} = \sqrt{\frac{6m_{\rm e}^{2}c^{4}e}{\sigma_{\rm T}B}} \simeq 1.1\,{\rm PeV} \; B_{\rm -3}^{-1/2}\;,
\label{E_max}
\end{equation}
where $B_{\rm -3}=B/(1$\,mG).

As we will see in Section~\ref{Results}, particles are accelerated to
high energies only if they are injected in a region of the TS close to
the equatorial plane. Therefore, to avoid computing uninteresting trajectories,
we introduce a free parameter 
$\mathcal{F}_{\rm inj}$, which we vary between roughly $5\%$ and 
$20\%$, according to the particular simulation, and we select for the injection region 
the range $|z| \leq \mathcal{F}_{\rm inj}z_{0}$. 
The normalization factor $Q_{0}$ depends on the fraction,
$\epsilon_{\rm acc,f}$, of particles injected at $|z| \leq
\mathcal{F}_{\rm inj} z_{0}$ that are accelerated to $E_{\rm d}
\geq f \times E_{\rm inj,d}$. We determine $\epsilon_{\rm acc,f}$
numerically. 

Let us assume that the equatorial region of the TS is approximately spherical with a
radius $r_{\rm TS}$, and that the region at $|z| \leq \mathcal{F}_{\rm
  inj} z_{0}$ in our planar 1D simulations corresponds to a
ring-shaped region of the TS whose half-width, as viewed from the
pulsar, subtends an angle $\Theta_{\rm inj} = \mathcal{F}_{\rm inj} z_{0}/r_{\rm TS}$. 
In this model, the angle 
$\Theta = z_{0}/r_{\rm TS}$ corresponds to that between the
rotation and magnetic axes of the pulsar. 
The angular dependence of the wind power, ${\rm d} L_{\rm s.d.}/{\rm d}\Omega$, 
can be modeled as being proportional to
$\sin^n\theta$, where $\theta$ is the colatitude, and the index $n$
lies between~2 (when the magnetic and rotation axes are aligned) and 4
(when they are orthogonal) \citep{tchekhovskoyetal16}. The angular dependence of 
the particle component, however, is not well constrained. Here, we assume it
has the same functional form, so that the rate at which electrons (or
positrons) are injected at $|z| \leq \mathcal{F}_{\rm inj} z_{0}$ is
\begin{eqnarray}
\dot{N}_{\rm \pm,inj} &=& 2\pi\int_{\pi/2-\Theta_{\rm inj}}^{\pi/2+\Theta_{\rm inj}} {\rm d}\theta\sin\theta 
\left({\rm d} L_{\rm s.d.}/{\rm d}\Omega\right)/E_{\rm inj,d}\\
&\approx& 
4\pi\left({\rm d} L_{\rm s.d.}/{\rm d}\Omega\right)_{\theta=\pi/2}
\mathcal{F}_{\rm inj} z_0/\left(r_{\rm TS}E_{\rm inj,d}\right)
\\
\noalign{\hbox{and $Q_{0}$ of Eq.~(\ref{defineq0}) is}}
Q_{0} &=& 
\frac{(\alpha_{\rm e}+1)\epsilon_{\rm acc,f}\dot{N}_{\rm \pm,inj}}
{(\gamma_{\max}^{\alpha_{\rm e}+1}-\gamma_{\min}^{\alpha_{\rm
    e}+1})} \times
\left\{ \begin{array}{ll}
1 & \textrm{for\ }n=0\\
1.5 & \textrm{for\ }n=2\;,\\
1.9 & \textrm{for\ }n=4\\
\end{array}\right.
\end{eqnarray}
where $\gamma_{\min} = f E_{\rm inj,d}/m_{\rm e}c^{2} $
and $\gamma_{\max} = E_{\max}/m_{\rm e}c^{2}$.

For convenience, we summarize here the main parameters of our simulations:
\begin{itemize}
\item[$z_0$:]
The height of the \lq\lq striped\rq\rq\ wind region at the TS.
\item[$\Theta$:]
The angle between the rotation and magnetic axes of the pulsar,
$\Theta=z_0/r_{\rm TS}$.
\item[$z_{\rm crit}$:]
The height at which the gyroradius of an injected particle in the large-scale magnetic field
(\ref{B_d}) equals its height above the equator, as defined in Eq.~(\ref{z_crit}) (independent of the 
level of turbulence).
\item[$\eta_{\rm crit}$:]
The ratio of the turbulent field to the large-scale field at height $z_{\rm crit}$.
\item[$z_{\rm w}$:]
The approximate height of the injection region of the TS that leads to effective acceleration, as 
estimated from Fig.~\ref{Spectra_Bands}. For $\eta_{\rm crit}<1$ (i.e., weak turbulence), 
$z_{\rm w}\sim\,$a~few$\,\times z_{\rm crit}$, but increases with the level of turbulence.
\item[$\mathcal{F}_{\rm inj}$:]
The height of the region of the TS at which particles are injected in the simulations, divided by $z_0$.
\item[$\epsilon_{\rm acc,f}$:]
The fraction of injected particles whose energy is boosted by at least a factor $f$, i.e., those 
accelerated to energy $>f\times E_{\rm inj,d}$, as determined from the simulations.
\end{itemize}

\section{Results}
\label{Results}

\subsection{Trajectories of electrons and positrons}
\label{Individual_Traj}

\begin{figure*}
  \centerline{\includegraphics[width=0.49\textwidth]{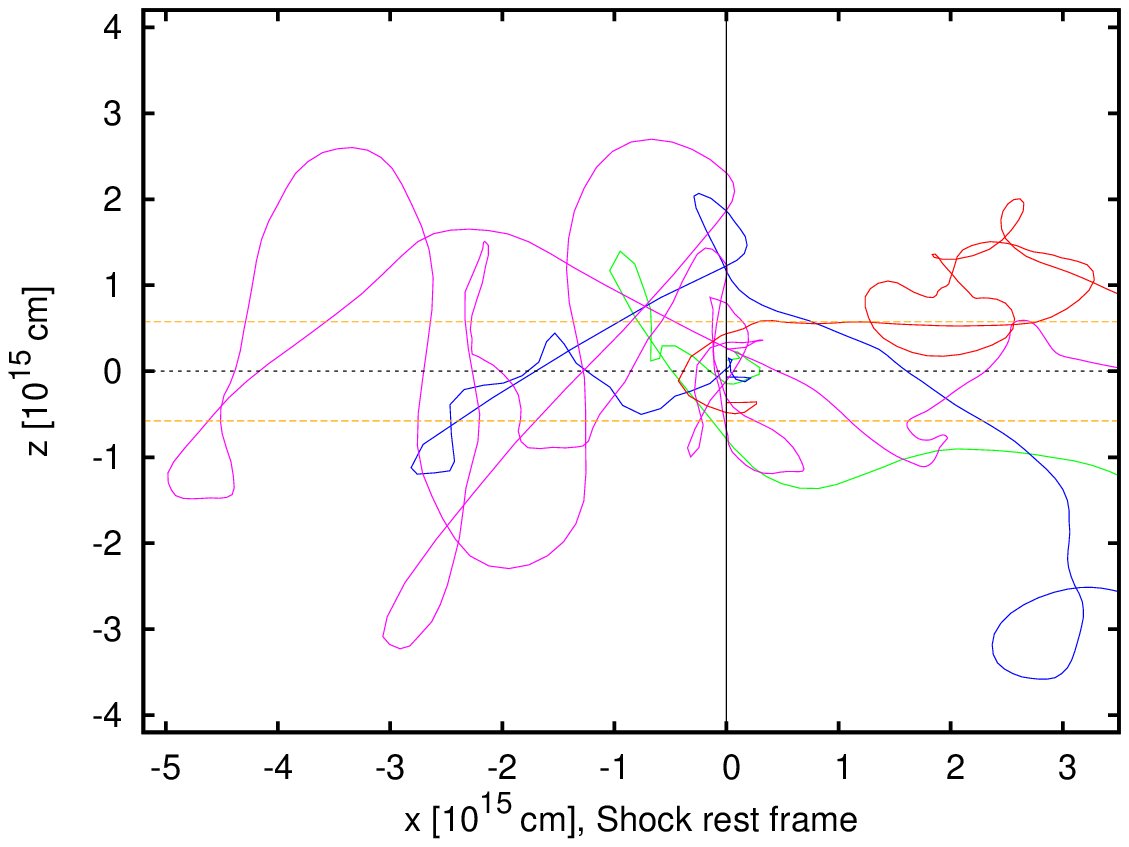}
              \hfil
              \includegraphics[width=0.49\textwidth]{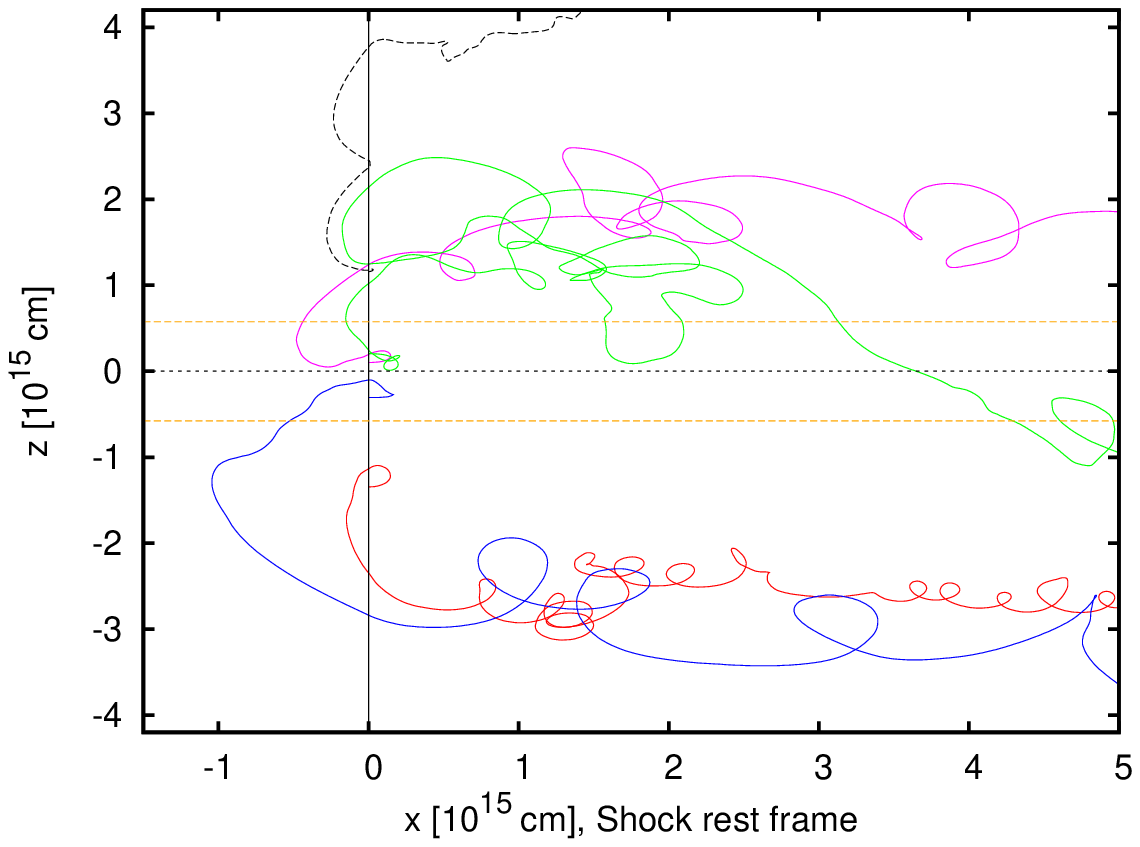}
              }
  \centerline{\includegraphics[width=0.49\textwidth]{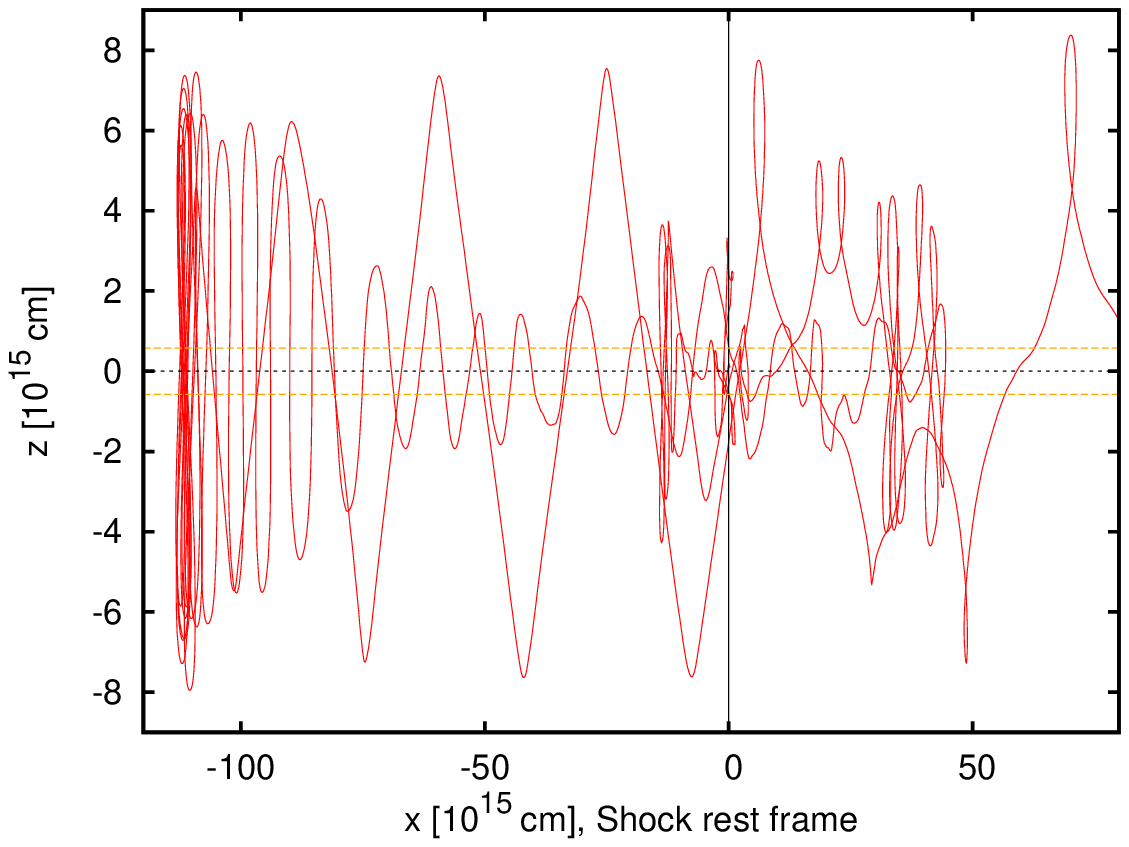}
              \hfil
              \includegraphics[width=0.49\textwidth]{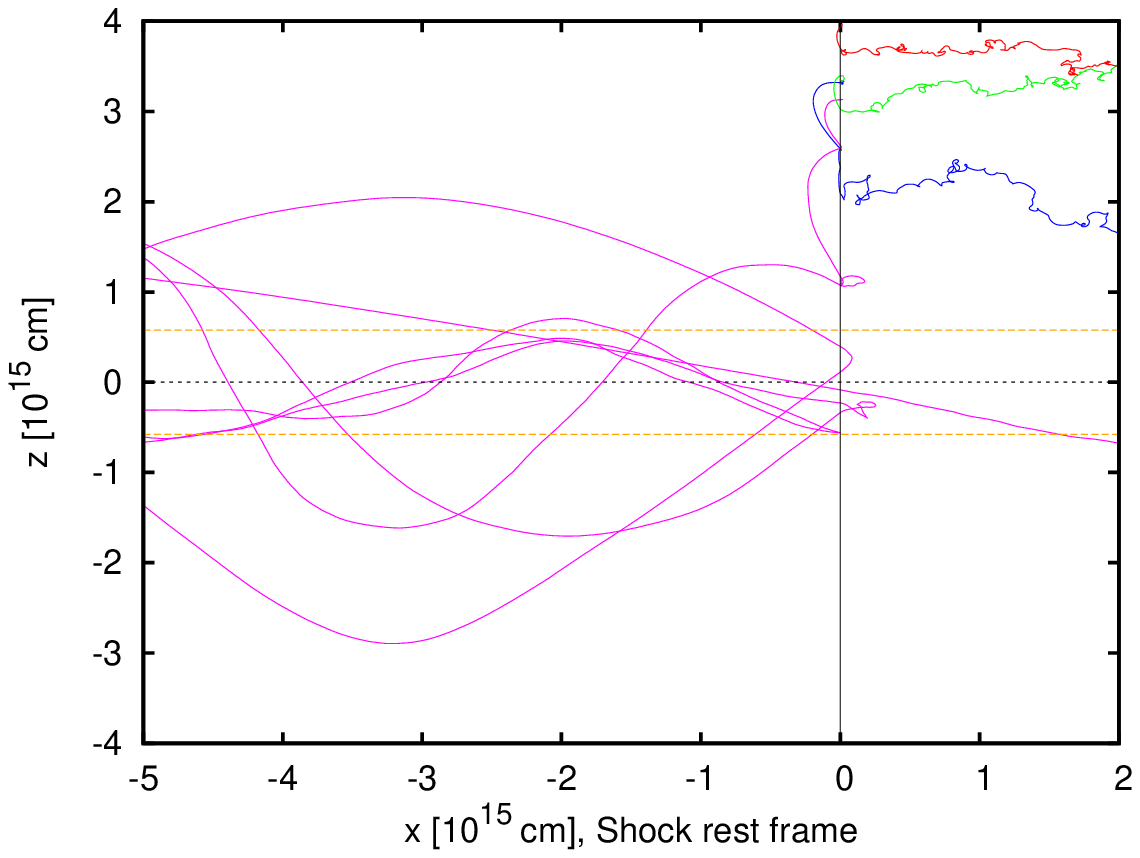}
              }
              \caption{Upper row and lower left panel: trajectories of
                electrons (left panels) and positrons (upper right
                panel) injected at $|z|/z_{0} \leq 0.015$, for $\delta
                B_{\rm d} = 30\,\mu$G (solid curves in the three
                panels) or $\delta B_{\rm d} = 400\,\mu$G (black
                dashed curve in the upper right panel). Lower right
                panel: trajectories of electrons injected at $z/z_{0}
                > 0.015$, and for $\delta B_{\rm d} = 400\,\mu$G. In
                all four panels, trajectories are plotted in the SRF
                and projected onto $(x,z)$. The parameters are $z_{0}=10^{17}$\,cm,
                $B_{\rm d,0}=1$\,mG, and $E_{\rm inj,d}=1$\,TeV. The
                vertical black solid lines at $x=0$ denote the shock
                position, the horizontal black dotted lines the
                equatorial plane ($z=0$), and the orange dashed lines 
                the critical distance $\pm z_{\rm crit}$ from the equatorial plane.}
\label{Traj_Ele_Pos}
\end{figure*}

\begin{figure}
  \includegraphics[width=0.49\textwidth]{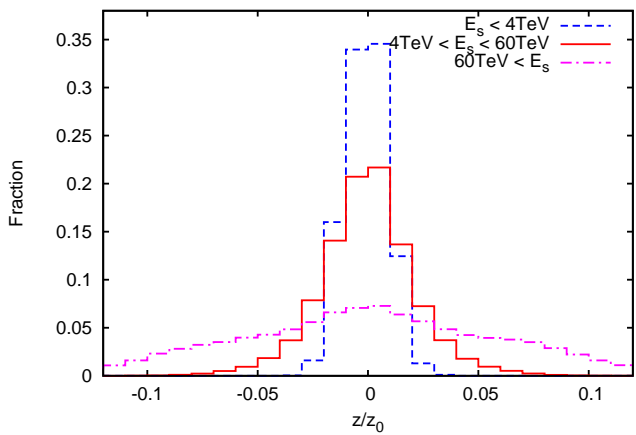}
  \caption{Distributions of normalized shock crossing altitudes, $z/z_{0}$, for electrons with energies $E_{\rm s} < 4$\,TeV (dashed blue line), $4\,{\rm TeV} < E_{\rm s} < 60$\,TeV (solid red line), and $E_{\rm s} > 60$\,TeV (magenta dash-dotted line) at the time of shock crossing. Electrons are injected at $|z|/z_{0} \leq 0.015$, $\delta B_{\rm d} = 30\,\mu$G, and the other parameters are set to the same values as in Fig.~\ref{Traj_Ele_Pos}.}
\label{Distribution_Z_30muG}
\end{figure}

First, we examine particle trajectories in the region of the TS that
is close to the equatorial plane, in the sense that
$\left|z\right|\lesssim z_{\rm crit}$.  (We show below that this
region is the most favorable for electron acceleration.)  In the
upper row of Fig.~\ref{Traj_Ele_Pos}, several trajectories in the SRF
are plotted for electrons (upper left panel) and positrons (upper
right panel) injected at the TS at $|z|/z_{0} \leq 0.015$, with
$z_{0}=10^{17}$\,cm. All other parameters are set to the values
discussed in \S~\ref{Model}. In particular, the injection energy is
$E_{\rm inj,d}=1$\,TeV, the magnetic field at $z_{0}$ is 1\,mG 
($B_{\rm d,0,-3}=1$), and the pulsar polarity is such that $B_{\rm
  d,0} > 0$. 
In the following, we refer to \lq\lq electrons\rq\rq\ and \lq\lq positrons\rq\rq\ 
for this pulsar polarity. For the opposite polarity, the situation
for electrons and positrons is inverted. 
The four solid lines (magenta, red, green, and blue)
in both panels represent typical particle trajectories, calculated for
$\delta B_{\rm d} = 30\,\mu$G and projected onto the $(x,z)$
plane. The level of turbulence, 
$\delta B_{\rm d}/B_{\rm d}=3/\left(z/10^{15}\,\textrm{cm}\right)$, 
at $z=z_{\rm crit}$ is, therefore,
$\eta_{\rm crit}\sim 5$. 
We show only examples of 
particles which return to the shock and enter the upstream region. In
the simulations, most injected particles ($\sim 90$\,\%) 
escape downstream without experiencing acceleration. 
The upstream region is on the left-hand side of the
panels, at $x<0$, and the downstream is on the right-hand side, at
$x>0$. The shock position is denoted by a thin vertical black line at
$x=0$, and the equatorial plane is marked by a dotted black line
at $z=0$. By comparing the two upper panels of
Fig.~\ref{Traj_Ele_Pos}, one can clearly see that electrons and
positrons behave differently. The drift-like motion imposed on crossing and recrossing the 
shock pushes positrons away from the equatorial plane, i.e. their $|z|$ tends to increase
with time, whereas electrons are pushed towards $z=0$ and remain on
orbits close to, or around, the equatorial plane. Despite the perturbations introduced
by the turbulent field, several of these
electrons spend time on trajectories that closely resemble Speiser orbits, 
such as the magenta
trajectory at $x<0$, in the upper left panel.
The fact that shock-drift systematically focuses the electrons into the
equatorial plane has a positive impact on their acceleration:
electrons tend to re-enter the downstream in regions with larger
turbulence levels $\delta B_{\rm d}/B_{\rm d}$, and thence have a
non-negligible probability to be scattered back into the upstream and
continue to gain energy via the first-order Fermi mechanism. Indeed,
one can see that the electrons plotted in the upper left panel cross
and re-cross the TS several times. The lower left panel shows the
trajectory of another electron accelerated to high energy. 
One can see that this electron spends
most of its time on Speiser orbits, although it spends some time on a
drift orbit, cf.~the two loops in the downstream at $x\approx
(1.5-2.5)\times 10^{16}$\,cm and $z\approx (3-5)\times
10^{15}$\,cm. The orbits appear irregular because of particle scattering
induced by the turbulent magnetic fields. We confirm that accelerated
electrons remain focused around the equatorial plane by plotting in
Fig.~\ref{Distribution_Z_30muG} the distribution of the normalized
shock crossing altitudes $z/z_{0}$ of electrons injected at $|z|/z_{0}
\leq 0.015$. In total, $5\times 10^{6}$ particles are
injected. We again use $\delta B_{\rm d} = 30\,\mu$G in this example, and
verified that the results are not significantly different for $\delta
B_{\rm d} = 400\,\mu$G. Three energy bands are shown, see the key in
the figure. One can clearly see that the electrons cross and re-cross
the TS in a small region around $z=0$, with a typical width of a few
percent of $z_{0}$. Even though the size of this region increases with
electron energy $E_{\rm s}$ (measured in the SRF), 
this is only due to the increase of the
particle gyroradius. We checked that electrons always remain well
confined and focused around $z=0$, even at the highest
energies.

In contrast, the situation for positrons is less favorable. As
can be seen in the upper right panel of Fig.~\ref{Traj_Ele_Pos}, those
that cross the TS and enter the upstream at $z=z_{1}$ 
re-enter the downstream at $|z|>|z_{1}|$. 
This is clearly visible for the red and blue 
trajectories at $z<0$. This forces the positrons to re-enter the 
downstream in regions where turbulence levels are lower. They are then 
more likely to be advected away from the shock due to the stronger
toroidal field at larger $|z|$, and this shuts down the first-order Fermi mechanism. 
Out of the four plotted positron trajectories, three of
them complete only one cycle (i.e., downstream $\rightarrow$ upstream
$\rightarrow$ downstream), and only one performs two (green trajectory). 
Increasing the strength of the turbulence
in the downstream increases the probability for positrons to complete
more cycles: the dashed black line shows a positron trajectory
for $\delta B_{\rm d} = 400\,\mu$G, which completes two cycles. However, even
in this case, acceleration quickly stops once the shock-induced drift pushes
the particle to larger $|z|$ where the turbulence levels are
smaller. One can see that this particle is advected in the downstream
at $z\simeq 4 \times 10^{15}$\,cm. Acceleration again stops more
quickly than for electrons. The orange dashed lines in
Fig.~\ref{Traj_Ele_Pos} show the altitudes where $z = \pm z_{\rm
  crit}$. For these parameter values, $z_{\rm crit} \simeq
0.0058\,z_{0}$. It is interesting to note that in the downstream, the
${\bf \nabla}B$-drift is strongest around $|z|\approx (1-3) \times
z_{\rm crit}$, and is directed towards the shock for positrons, both
at $z>0$ and $z<0$. In other words, the ${\bf \nabla}B$-drift helps
the positrons injected in these regions to fight against advection,
and it increases their chances of entering the upstream for their
first cycle (e.g., the first half-gyration in the downstream for
the red and blue trajectories in the upper right panel in
Fig.~\ref{Traj_Ele_Pos}). Ultimately, however, this is to no avail,
because of the effect of shock-drift during the first cycle.

In the lower right panel of Fig.~\ref{Traj_Ele_Pos}, we show 
the trajectories of four
electrons injected further from the equatorial
plane, at $3 \times
10^{15}\,{\rm cm} < z < 4 \times 10^{15}$\,cm, and take $\delta B_{\rm
  d} = 400\,\mu$G, the other parameter values remaining unchanged. 
It is apparent that shock-drift pushes all these electrons closer to
$z=0$. Because of the lower turbulence levels in the downstream at
these larger $|z|$, the probability for a particle in the downstream
to be scattered back into the upstream is smaller, and out of the four
plotted trajectories, only one of them reaches the equatorial plane
(the magenta line). The other three are advected away downstream after 
only one or a very few cycles. For example, the green
trajectory completes one excursion into the upstream, whereas the blue one
completes three. These electrons do gain some energy, thanks to
the first-order Fermi effect and the shock-induced drift. However, the electron with the
magenta trajectory gains significantly more energy than the others,
because it reaches the equatorial plane region, which is the
most favorable one for particle acceleration. Once an electron enters 
this region, it remains on Speiser orbits, as do those 
injected at $|z|/z_{0} < 0.015$; see the oscillations between $z>0$
and $z<0$ in the upstream.

\subsection{Particle spectrum close to the equatorial plane}
\label{Results_Equatorial_Plane}

\begin{figure*}
  \centerline{\includegraphics[width=0.49\textwidth]{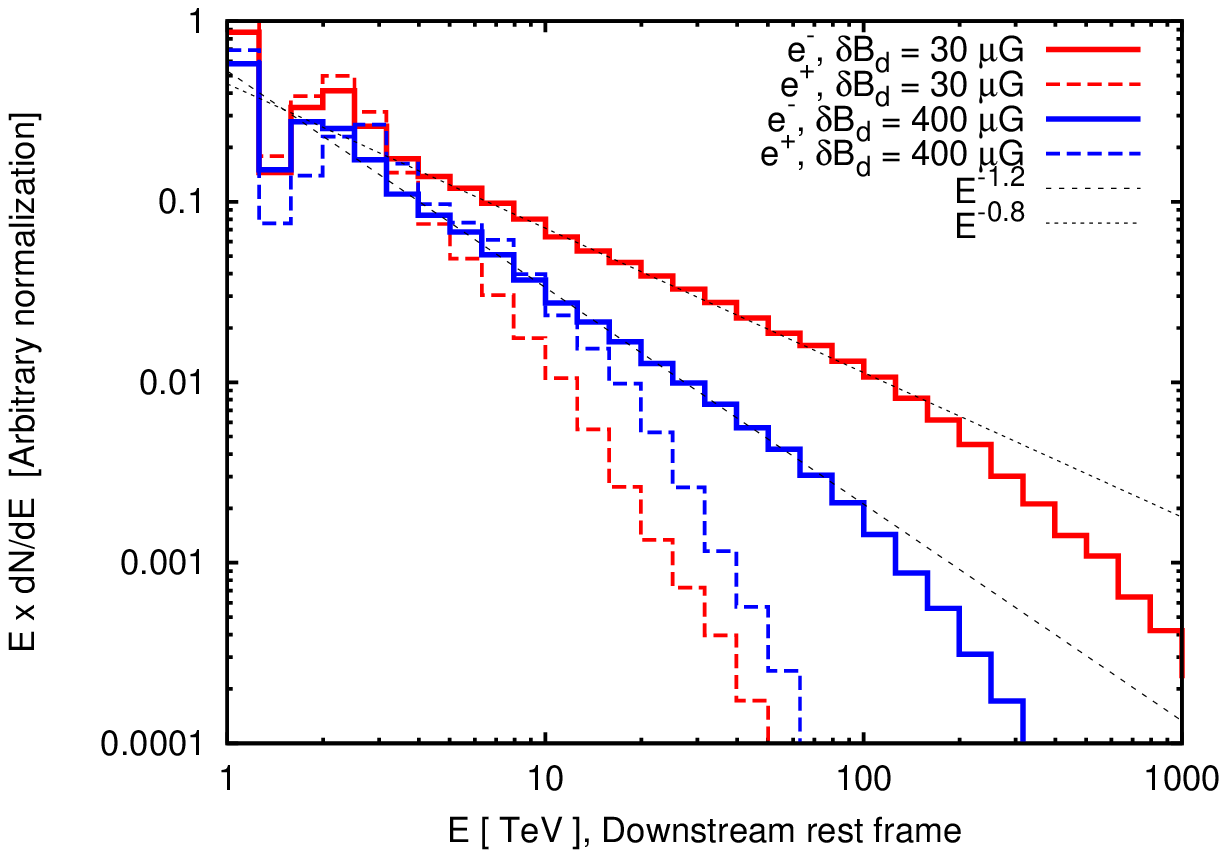}
              \hfil
              \includegraphics[width=0.49\textwidth]{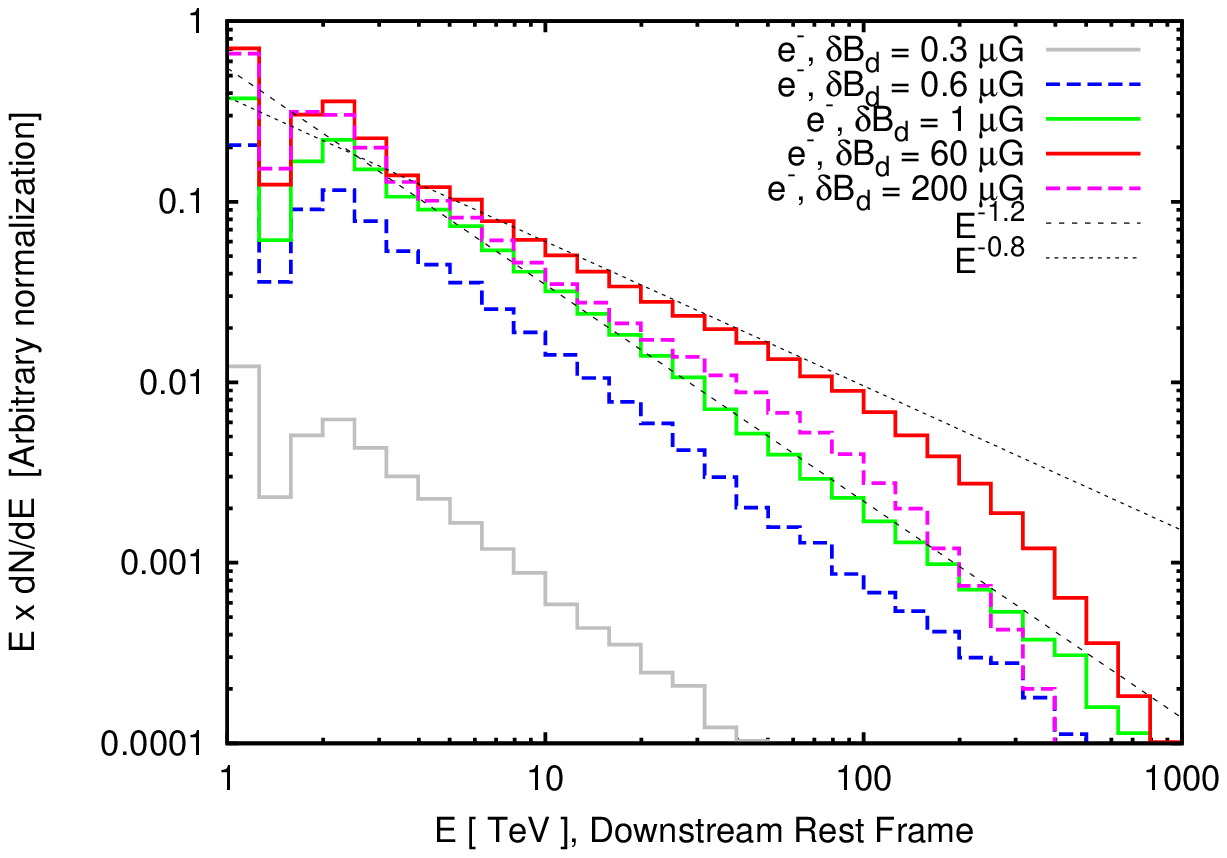}
              }
              \caption{Left panel: spectra $E_{\rm d} \times
                dN/dE_{\rm d}$ of electrons (thick solid lines) and
                positrons (dashed lines) in the DRF, for $\delta
                B_{\rm d} = 30\,\mu$G (red lines, corresponds to
                $\eta_{\rm crit} = 5.2$) and $\delta B_{\rm d} =
                400\,\mu$G (blue lines, $\eta_{\rm crit} = 69$); Right
                panel: spectra $E_{\rm d} \times dN/dE_{\rm d}$ of
                electrons for $\delta B_{\rm d} =
                0.3,\,0.6,\,1,\,60,\,200\,\mu$G (i.e. $\eta_{\rm crit}
                = 0.052$, 0.10, 0.17, 10, 35). See the key for the
                corresponding line types and colors. In both panels,
                $z_{0}=10^{17}$\,cm, $B_{\rm d,0}=1$\,mG, and the
                particles are injected at $|z|/z_{\rm crit} \leq 2.6$
                (i.e. $|z|/z_{0} \leq 0.015$) with $E_{\rm
                  inj,d}=1$\,TeV. For reference, the thin black dashed
                (dotted) lines show power-laws $\propto E_{\rm
                  d}^{-1.2}$ (respectively $\propto E_{\rm d}^{-0.8}$).}
\label{Spectra_Ele_Pos_Equator}
\end{figure*}

We calculate now the energy spectrum of the particles injected and
accelerated in the equatorial region of the TS. The injection region
where particles are most likely to reach high energies is typically
within a few $z_{\rm crit}$ from the equatorial plane. We denote the 
height of this region by $z_{\rm w}$, and find
(cf.\ Sect.~\ref{Whole_TS}) $z_{\rm w}\approx 5\times10^{14}\,$cm 
for $z_{0} = 10^{16}$\,cm, $z_{\rm w}\approx 1.5\times 10^{15}\,$cm for $z_{0} = 10^{17}$\,cm, 
and $z_{\rm w} \approx 3.6\times 10^{15}\,$cm for $z_{0} = 6 \times 10^{17}$\,cm. 
In the following, we consider the latter two cases. For
each tested set of parameters, we inject $5\times 10^{6}$ particles 
at points equally spaced in $z$ in
this region and construct the spectrum by recording the particle
energy in the DRF at each shock crossing. 
Since particles do not change their energy whilst in the DRF, 
the steady-state spectrum at the shock, averaged over all injection points, 
is identical to the spectrum of particles at $x=d$, where they are considered to 
have escaped. However, much better statistics are achieved by binning the spectrum
at each shock crossing, rather than only at escape. We plot
the steady-state spectrum in the DRF, without taking 
into account the particles that have been advected in the
downstream without being accelerated, i.e., the spectra shown hereafter
refer to particles that have performed at least one cycle.

In Fig.~\ref{Spectra_Ele_Pos_Equator} (left panel), we plot the
spectra $E_{\rm d} \times dN/dE_{\rm d}$ of electrons (thick solid
lines) and positrons (dashed lines) injected at $|z|/z_{0} \leq 0.015$
for $z_{0} = 10^{17}$\,cm (i.e. $|z|/z_{\rm crit} \leq 2.6$), and for two levels of turbulence in the
downstream: $\delta B_{\rm d} = 30\,\mu$G (red lines) and 
$\delta B_{\rm d} = 400\,\mu$G (blue lines), corresponding to a level of
turbulence at $z=z_{\rm crit}$ of $\eta_{\rm crit} = 5.2$ and $\eta_{\rm crit} = 69$, respectively.  The
positron spectra are much softer than the electron spectra, 
even in the most favorable case of strong
turbulence in the downstream, $\delta B_{\rm d} = 400\,\mu$G. This
confirms the trend found in the previous subsection: only electrons
are efficiently accelerated, whereas positrons are expelled from the
acceleration region before they can reach high energies. The electron
spectra in Fig.~\ref{Spectra_Ele_Pos_Equator} extend to $E_{\rm d}
\sim (100 - 300)$\,TeV. These high-energy cutoffs are an artifact. 
They occur at the energy at which the electron
gyroradius equals the maximum size $L_{\max}$ of the grid on
which the turbulent field is defined, above which the scattering is strongly 
suppressed. 
In contrast, the cutoffs in the positron spectra are physical, because 
they appear below that energy. We demonstrate these points in the Appendix by repeating the 
calculations of Fig.~\ref{Spectra_Ele_Pos_Equator} (left panel) with a smaller value 
of $L_{\max}$, and a reduced grid size.
Below the $\sim (100 - 300)$\,TeV cutoff and above $E_{\rm d}\gtrsim 4$\,TeV,
i.e., above a few times the injection energy,
the electron spectra are well described by power-laws. To guide the eye, we plot
two power-laws: one $\propto E_{\rm d}^{-1.2}$ (thin dashed black
line), and the other $\propto E_{\rm d}^{-0.8}$ (thin dotted black
line). One can clearly see that the electron spectral index depends on
$\delta B_{\rm d}$, being $\alpha_{\rm e} \simeq -1.8$ for
$\delta B_{\rm d} = 30\,\mu$G, and $\alpha_{\rm e} \simeq -2.2$ for
$\delta B_{\rm d} = 400\,\mu$G. We note that the latter value of
$\alpha_{\rm e}$ is compatible with the index expected for particles
accelerated at a relativistic shock with pure scattering and no
large-scale magnetic field \citep{Achterberg2001}. 
In Fig.~\ref{Spectra_Ele_Pos_Equator} (right panel), we plot electron
spectra for a wider range of values of $\delta B_{\rm d}$: $\delta
B_{\rm d} = 0.3\,\mu$G (solid grey line), $0.6\,\mu$G (dashed blue),
$1\,\mu$G (solid green), $60\,\mu$G (solid red), and $200\,\mu$G
(dashed magenta), corresponding to levels of turbulence at $z_{\rm crit}$ of 
$\eta_{\rm crit} = 0.052$, 0.10, 0.17, 10, and 35. 
The electron spectrum is seen to be slightly softer
than $E_{\rm d}^{-2.2}$ for $\delta B_{\rm d} = (0.3$ -- $1)\,\mu$G. It
hardens to $dN/dE_{\rm d} \propto E_{\rm d}^{-1.8}$ for $\delta B_{\rm
  d} = 60\,\mu$G, and softens again for larger turbulence
levels: the dotted magenta line for $\delta
B_{\rm d} = 200\,\mu$G is compatible with an index $-2.2 < \alpha_{\rm
  e} < -1.8$. All curves are normalized to the same (arbitrary) level,
which shows that, for low levels of turbulence $\delta B_{\rm d} <
1\,\mu$G, a smaller fraction of the injected electrons are
accelerated. 

\begin{figure*}
  \centerline{\includegraphics[width=0.49\textwidth]{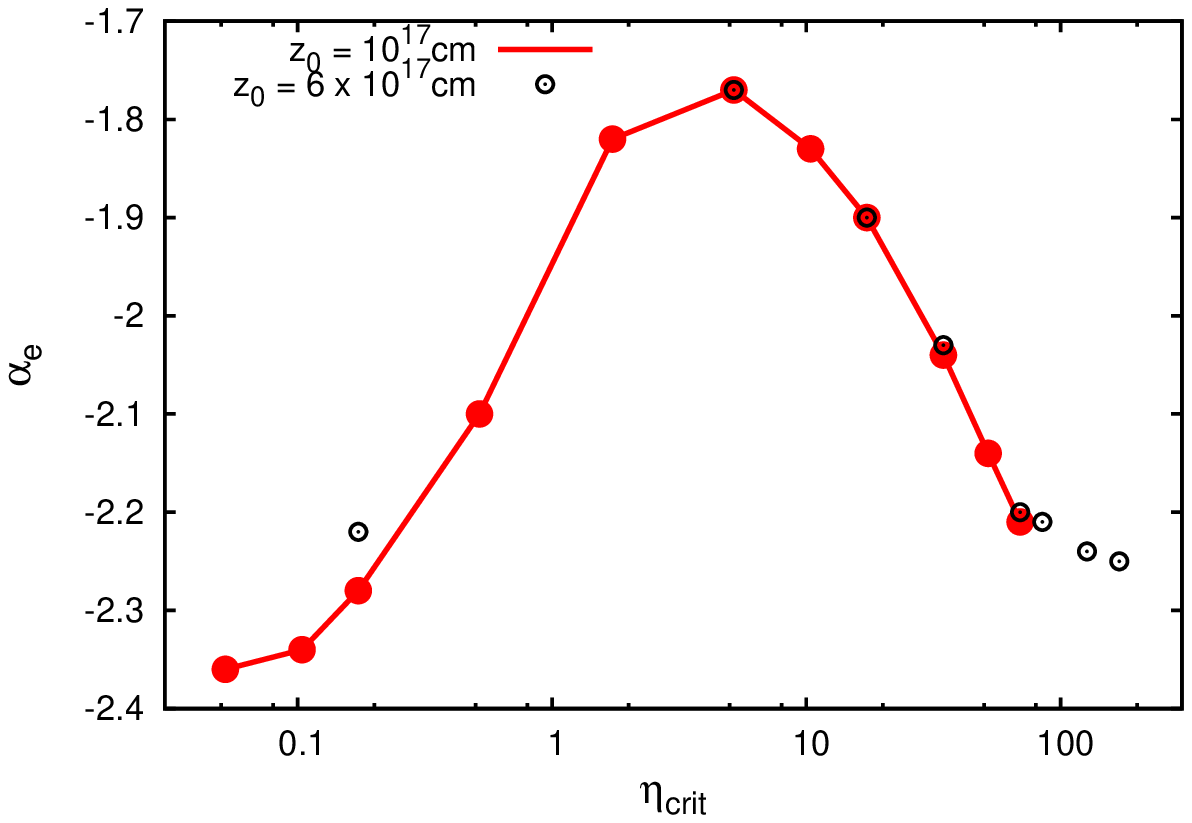}
              \hfil
              \includegraphics[width=0.49\textwidth]{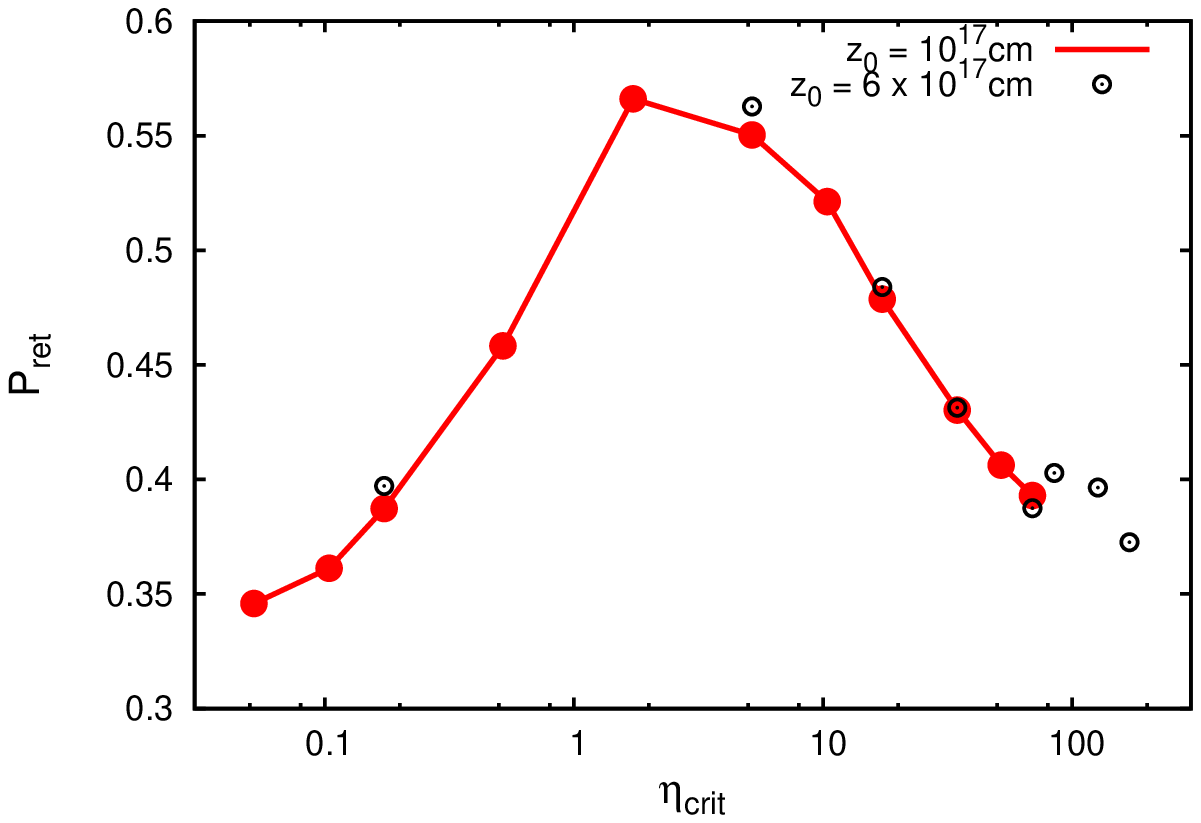}
              }
  \caption{Left panel: electron spectral index $\alpha_{\rm e}$ as a function of $\eta_{\rm crit}$ (fits on the interval $7\,{\rm TeV} \leq E_{\rm d} \leq 80$\,TeV). Right panel: return probability $\mathcal{P}_{\rm ret}$ as a function of $\eta_{\rm crit}$, for electrons with $7\,{\rm TeV} \leq E_{\rm d} \leq 80$\,TeV. On both panels, solid red lines are for $z_{0} = 10^{17}$\,cm, and open black circles for $z_{0} = 6 \times 10^{17}$\,cm. $B_{\rm d,0}=1$\,mG, and the electrons are injected at $|z|/z_{\rm crit} \leq 2.6$ (i.e. $|z|/z_{0} \leq 0.015 / \sqrt{z_{0}/10^{17}\,{\rm cm}}$) with $E_{\rm inj,d}=1$\,TeV.}
\label{Gamma_e_Equator}
\end{figure*}

\begin{deluxetable*}{ccccccc}
\tablecolumns{7}
\tablecaption{Simulations with injection close to the equatorial plane
\label{Table_Equator}}
\tablehead{\colhead{$z_{0}/(10^{17}\,{\rm cm})$} & \colhead{$\delta B_{\rm d}/(1\,\mu{\rm G})$} & 
\colhead{Turbulence level} & \colhead{Electron index} & \colhead{Return probability} 
& \colhead{Gain per cycle} 
& \colhead{Fraction at $>3\,$TeV}
\\
&& 
\colhead{$\eta_{\rm crit}$} & \colhead{$\alpha_{\rm e}$} & \colhead{$\mathcal{P}_{\rm ret}$} 
& \colhead{$(\Delta E/E)_{\rm d}$} 
& \colhead{$\epsilon_{\rm acc,3}$} \vspace{0.03cm}}
\startdata 
1 & 0.3   & $5.2 \times 10^{-2}$ & $-2.36 \pm 0.03$ & 0.35 & 1.05 & $1.1 \times 10^{-3}$ \\
1 & 0.6   & 0.10   & $-2.34 \pm 0.02$ & 0.36 & 1.07 & $2.1 \times 10^{-2}$ \\
1 & 1     & 0.17   & $-2.28 \pm 0.02$ & 0.39 & 1.06 & $4.4 \times 10^{-2}$ \\
1 & 3     & 0.52   & $-2.10 \pm 0.02$ & 0.46 & 1.08 & $6.9 \times 10^{-2}$ \\
1 & 10    & 1.7    & $-1.82 \pm 0.03$ & 0.57 & 1.07 & $8.4 \times 10^{-2}$ \\
1 & 30    & 5.2    & $-1.77 \pm 0.01$ & 0.55 & 1.10 & $5.7 \times 10^{-2}$ \\
1 & 60    & 10     & $-1.83 \pm 0.01$ & 0.52 & 1.11 & $5.3 \times 10^{-2}$ \\
1 & 100   & 17     & $-1.90 \pm 0.01$ & 0.48 & 1.09 & $5.2 \times 10^{-2}$ \\
1 & 200   & 35     & $-2.04 \pm 0.01$ & 0.43 & 1.09 & $4.7 \times 10^{-2}$ \\
1 & 300   & 52     & $-2.14 \pm 0.02$ & 0.41 & 1.09 & $4.8 \times 10^{-2}$ \\
1 & 400   & 69     & $-2.21 \pm 0.01$ & 0.39 & 1.08 & $4.5 \times 10^{-2}$ \\
6 & 0.41  & 0.17   & $-2.22 \pm 0.01$ & 0.40 & 1.05 & $3.2 \times 10^{-2}$ \\
6 & 12    & 5.2    & $-1.77 \pm 0.01$ & 0.56 & 1.09 & $6.3 \times 10^{-2}$ \\
6 & 41    & 17     & $-1.90 \pm 0.01$ & 0.48 & 1.11 & $5.0 \times 10^{-2}$ \\
6 & 82    & 35     & $-2.03 \pm 0.02$ & 0.43 & 1.14 & $4.4 \times 10^{-2}$ \\
6 & 163   & 69     & $-2.20 \pm 0.03$ & 0.39 & 1.13 & $4.7 \times 10^{-2}$ \\
6 & 200   & 85     & $-2.21 \pm 0.03$ & 0.40 & 1.16 & $4.9 \times 10^{-2}$ \\
6 & 300   & $1.3 \times 10^{2}$ & $-2.24 \pm 0.04$ & 0.40 & 1.14 & $4.6 \times 10^{-2}$ \\
6 & 400   & $1.7 \times 10^{2}$ & $-2.25 \pm 0.05$ & 0.37 & 1.16 & $4.2 \times 10^{-2}$%
\tablecomments{Electrons are injected at $|z|/z_{\rm crit} \leq 2.6$ 
corresponding to $|z|/z_{0} \leq 0.015/\sqrt{z_{0}/(10^{17}\,{\rm cm})}$. The injection
energy is $E_{\rm inj,d}=1$\,TeV and the regular magnetic field at $z_0$
is $B_{\rm d,0} = 1$\,mG.}
\enddata 
\end{deluxetable*}

In Table~\ref{Table_Equator}, seventh column, we give the fraction, 
$\epsilon_{\rm acc,3}$, of injected electrons that are
accelerated to $E_{\rm d} \geq 3$\,TeV. For $\delta B_{\rm d} < 1\,\mu$G,
$\epsilon_{\rm acc,3}$ quickly drops, but otherwise remains in the range
$\simeq 4 - 8$\%.
In the fourth column of Table~\ref{Table_Equator}, we provide the
values of $\alpha_{\rm e}$ for $z_{0} = 10^{17}$\,cm and 
$\delta B_{\rm d}$ within the range $(0.3 - 400)\,\mu$G, and for 
$z_{0} = 6 \times 10^{17}$\,cm and $\delta B_{\rm d} = (0.41 - 400)\,\mu$G. 
The third column contains the corresponding values of $\eta_{\rm crit}$. 
The spectral indexes are calculated by fitting the 
electron spectra on the energy interval $7\,{\rm TeV} \leq E_{\rm d} \leq 80$\,TeV 
where they are well described by power-laws.

In Fig.~\ref{Gamma_e_Equator} (left panel), we plot $\alpha_{\rm e}$
versus $\eta_{\rm crit}$. The 
red line and solid red dots are for $z_{0} = 10^{17}$\,cm, and the
open black circles are for $z_{0} = 6 \times 10^{17}$\,cm. The shape
of the red curve confirms the trend already noted in
Fig.~\ref{Spectra_Ele_Pos_Equator}. The spectrum is soft, with
$\alpha_{\rm e} \simeq -(2.3$ -- $2.2)$, at small ($\lesssim 1$) 
and large ($\gtrsim 30$) values of
$\eta_{\rm crit}$, i.e., small and large values of $\delta B_{\rm d}$. 
It hardens at intermediate values of $\eta_{\rm crit}$, and the index 
reaches its maximum of $\alpha_{\rm e} \simeq -1.8$ 
around $\eta_{\rm crit} \simeq {\rm a~few}$, 
i.e., when the turbulence level at $z_{\rm crit}$ is close to unity. 
The results for $\alpha_{\rm e}$ versus $\eta_{\rm crit}$ are
almost the same for both values of $z_{0}$, which suggests that 
$\alpha_{\rm e}$ is a function of $\eta_{\rm crit}$. We note that, at 
$B_{\rm d,0}$ fixed, 
$\eta_{\rm crit} \propto \delta B_{\rm d} \times \sqrt{z_{0}}$.

In the nonrelativistic theory of diffusive shock acceleration, the
spectral index $\alpha_{\rm e}$ is determined by the ratio of the
average return probability of electrons from downstream to upstream,
$\mathcal{P}_{\rm ret}$, and their average relative energy gain per
cycle, $\Delta E/E$ \citep{bell78}. The relativistic theory is more
complicated, since the (angular dependent) ratio of these quantities
must be convolved with the actual angular distribution of particles at
the shock.  Nevertheless, these quantities, separately averaged, give
a good intuitive guide to the mechanisms at work. In the fifth and
sixth columns of Table~\ref{Table_Equator}, we give the values of
$\mathcal{P}_{\rm ret}$ and $(\Delta E/E)_{\rm d}$ (i.e., $\Delta
E/E$ as measured in the DRF) respectively, for electrons with energies
$7\,{\rm TeV} \leq E_{\rm d} \leq 80$\,TeV. No clear trend emerges for
$(\Delta E/E)_{\rm d}$, and the results are compatible with $(\Delta
E/E)_{\rm d}$ being almost constant and $\simeq 1.1$. On the other
hand, $\mathcal{P}_{\rm ret}$ shows a strong variation with $\eta_{\rm crit}$. 
In Fig.~\ref{Gamma_e_Equator} (right panel), we plot
$\mathcal{P}_{\rm ret}$ versus $\eta_{\rm crit}$ for $z_{0} = 10^{17}$\,cm and 
$z_{0} = 6 \times 10^{17}$\,cm, with the same color code as in the left 
panel. The good match between the open black circles and the red curve 
shows that $\mathcal{P}_{\rm ret}$ is also a function of 
$\eta_{\rm crit}$. By comparing the 
left and the right panels in Fig.~\ref{Gamma_e_Equator}, one sees that 
$\alpha_{\rm e}$ and $\mathcal{P}_{\rm ret}$ are strongly
correlated. The return probability of electrons is maximal
($\mathcal{P}_{\rm ret} \approx 0.6$) at values of $\eta_{\rm crit}$
where the electron spectrum is hardest, and it is smaller
($\mathcal{P}_{\rm ret} \approx 0.35$ -- $0.4$) at values of $\eta_{\rm crit}$ 
where the electron spectrum is soft, $\alpha_{\rm e}
\approx -(2.3$ -- $2.2)$. This implies that the hard electron spectrum
found at $\eta_{\rm crit} \sim (1 - 10)$ is due to an increase 
in the return probability of the electrons 
from the downstream to the upstream at these turbulence levels. 
The reason is connected with the nature of the drift trajectories, 
combined with the fact that for $\eta\sim1$, the role of turbulence is 
significant on those sections of the orbit closer to the equatorial plane 
(low altitude, i.e., smaller $\left|z\right|$), and relatively unimportant
on those at higher altitude (larger $\left|z\right|$). 
Electron drift trajectories move away from the shock front 
($\dot{x}>0$) at low altitude, and towards it at high altitude. Since the turbulence 
predominantly scatters the low-altitude section, the net result is 
a reduction in $\dot{x}$, i.e., in the escape probability.

\subsection{Overall electron spectrum at the termination shock}
\label{Whole_TS}

\begin{figure*}
  \centerline{\includegraphics[width=0.32\textwidth]{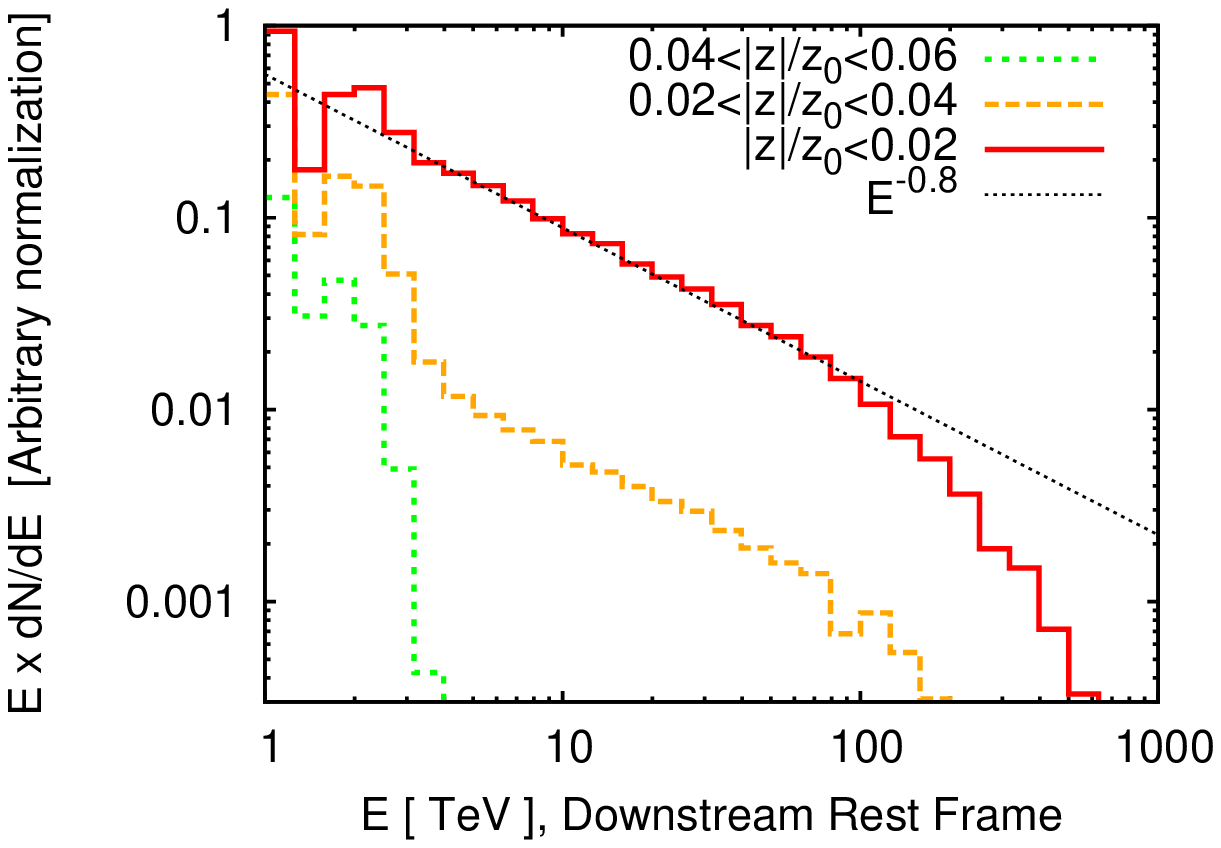}
              \hfil
              \includegraphics[width=0.32\textwidth]{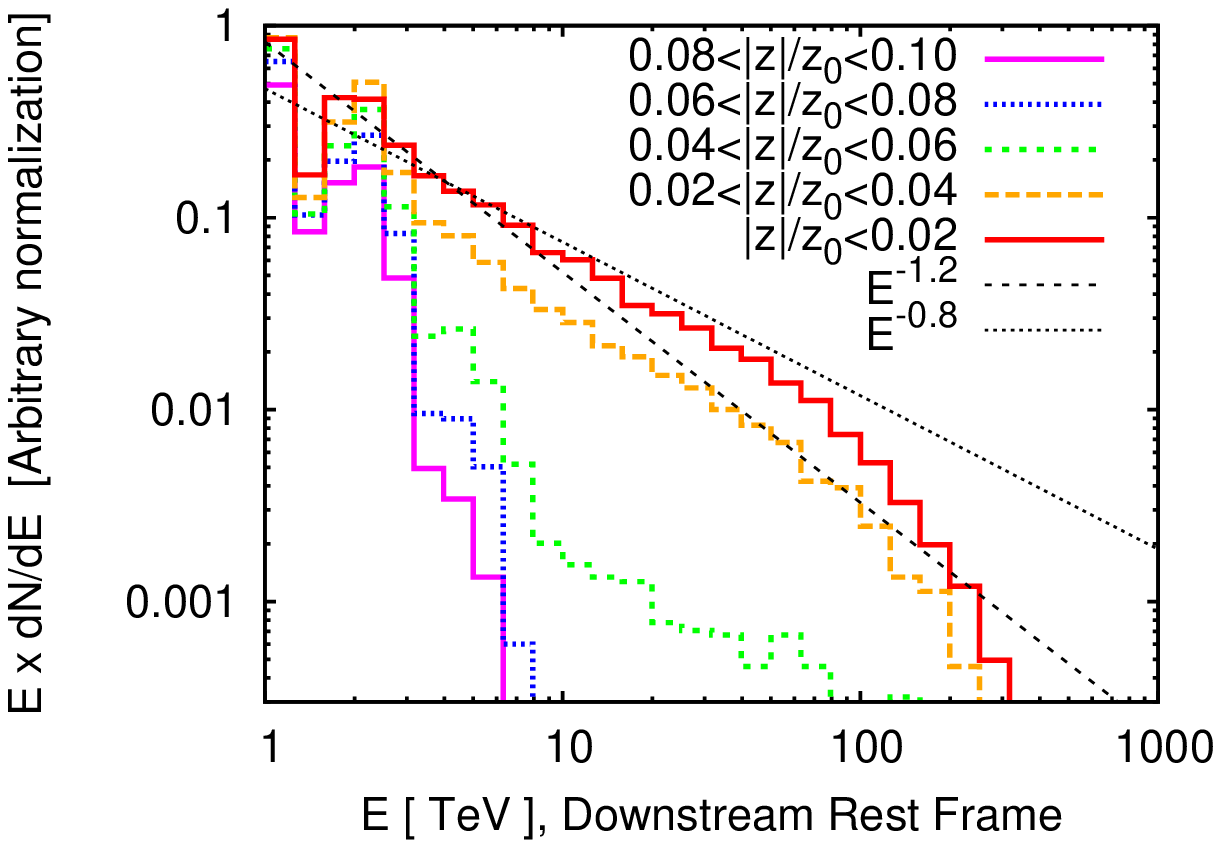}
              \hfil
              \includegraphics[width=0.32\textwidth]{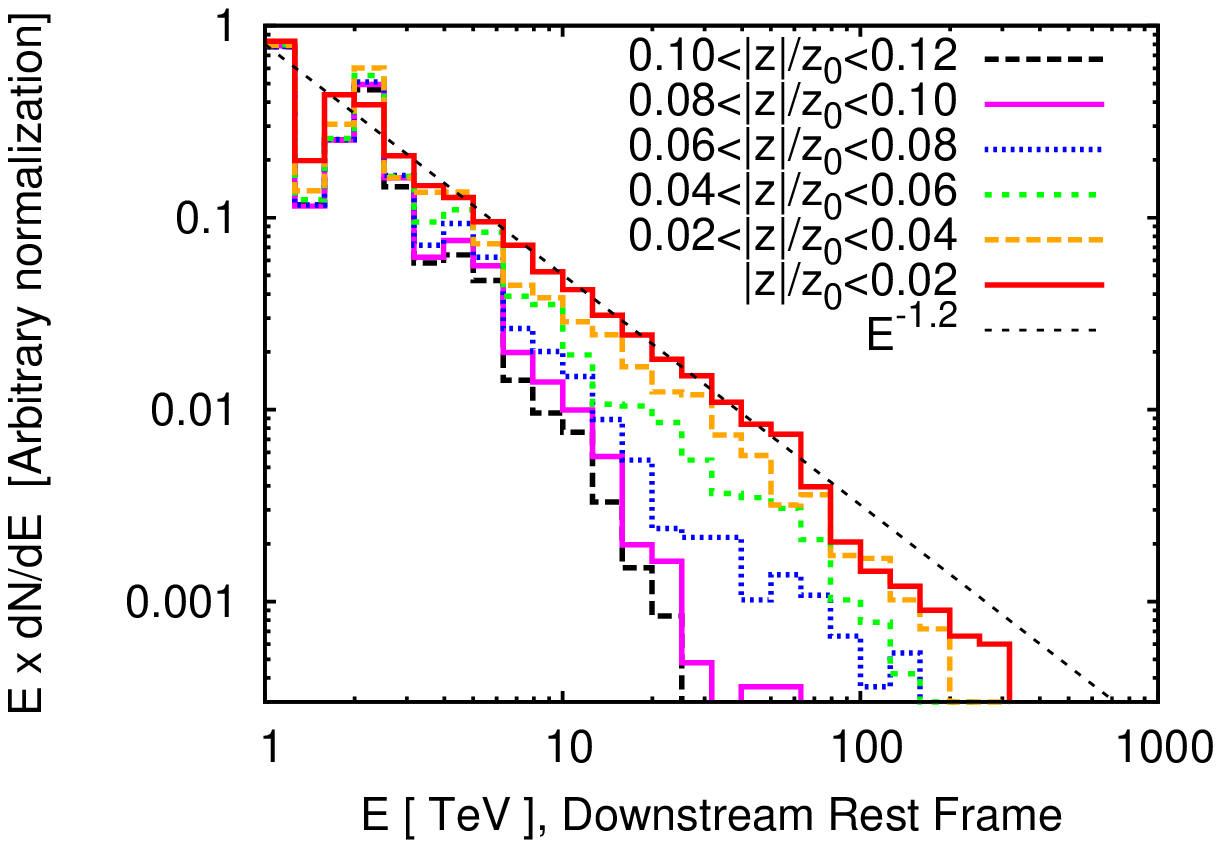}
              }
  \centerline{\includegraphics[width=0.32\textwidth]{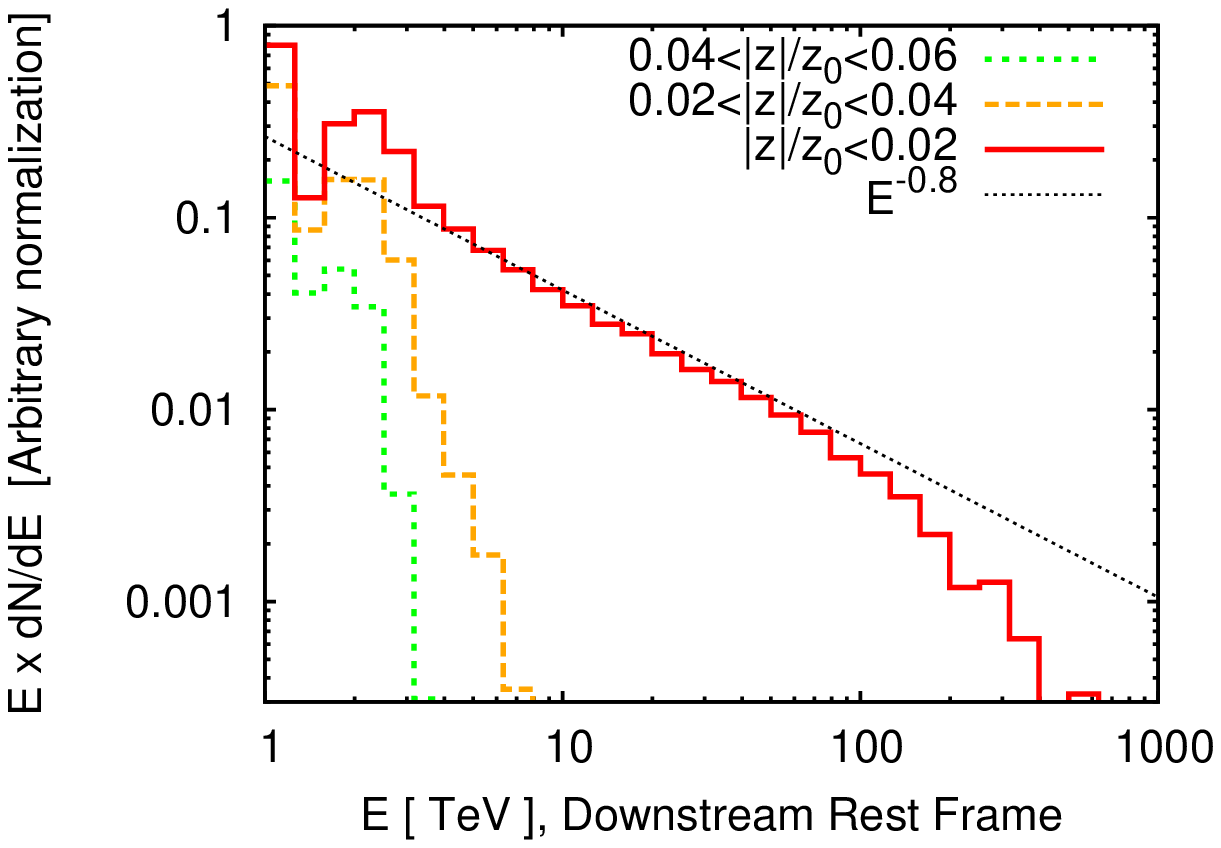}
              \hfil
              \includegraphics[width=0.32\textwidth]{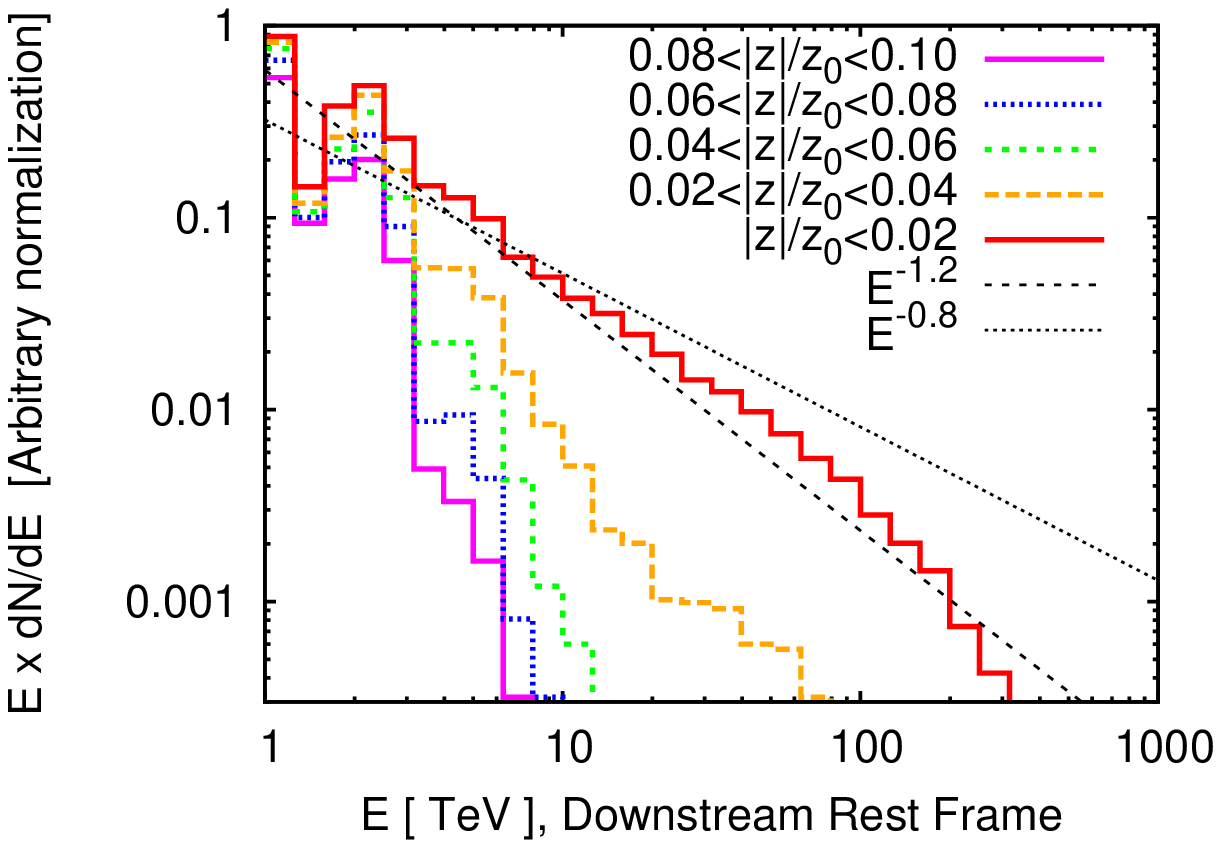}
              \hfil
              \includegraphics[width=0.32\textwidth]{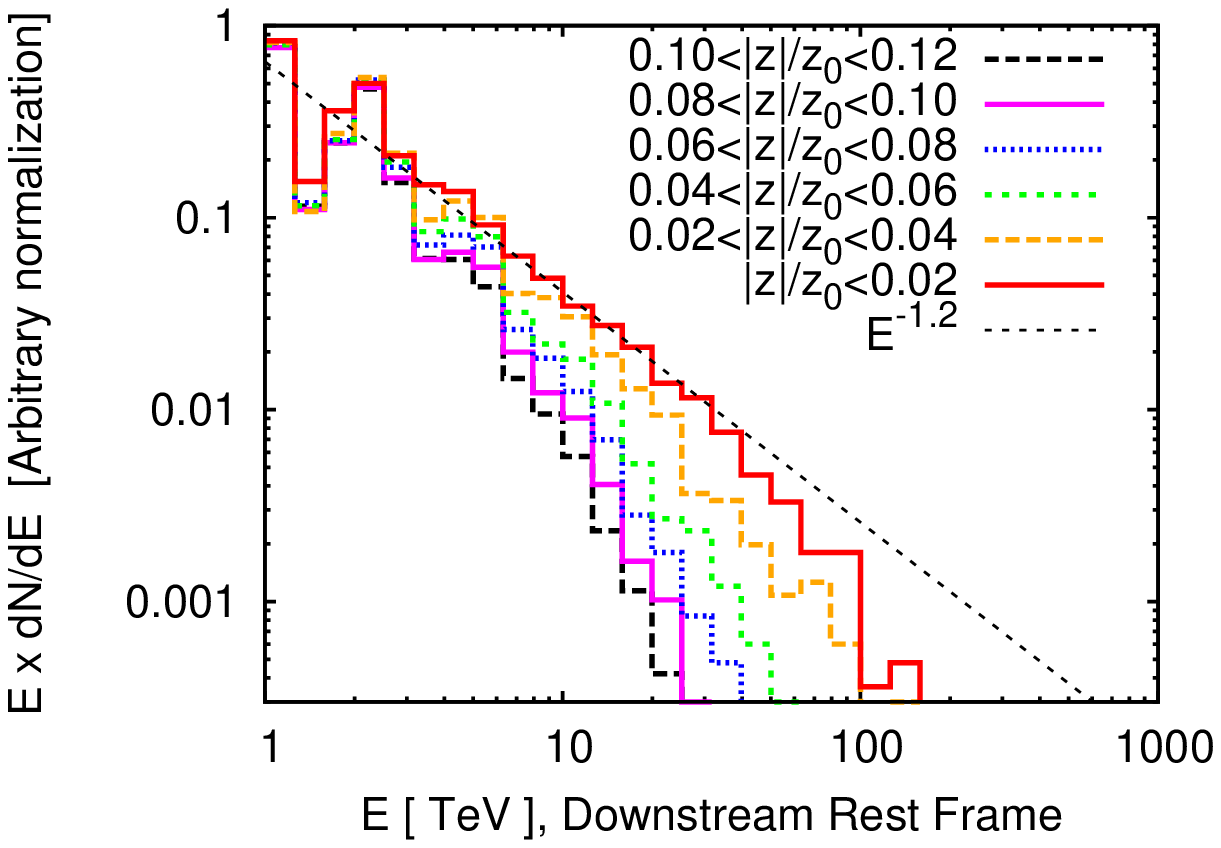}
              }
              \caption{Electron spectra $E_{\rm d} \times dN/dE_{\rm
                  d}$ in the DRF for injection close to the equatorial plane, 
$\left|z\right|/z_{0} \leq 0.02$ (solid red lines), and in five zones of increasing altitude above it:
$\zeta_i<\left|z\right|/z_0<\zeta_{i+1}$, with 
$\zeta_i=0.02\times i$, $i=1,\dots 5$, corresponding to 
$\xi_i<\left|z\right|/z_{\rm crit}<\xi_{i+1}$, with $\xi=\left(3.46,6.93,10.4,13.9,17.3,20.8\right)$
in the first row, where $z_0=10^{17}\,$cm, and with    
$\xi=\left(8.48,17.0,25.4,33.9,42.4,50.9\right)$ in the second row, where $z_0=6\times10^{17}\,$cm. 
In the first column $\delta B_{\rm d} = 30\,\mu$G, in the second $\delta B_{\rm d} =
100\,\mu$G, and in the third $\delta B_{\rm d} = 400\,\mu$G. 
The thin black dashed and dotted lines show power-laws $\propto E_{\rm d}^{-1.2}$ and $\propto E_{\rm d}^{-0.8}$, respectively.}
\label{Spectra_Bands}
\end{figure*}

We now investigate the acceleration, or lack thereof, of electrons
injected further away from the equatorial plane. In
Fig.~\ref{Spectra_Bands}, we plot the spectra $E_{\rm d} \times
dN/dE_{\rm d}$ of electrons injected at the TS in six different zones
of equal area, located at successively increasing distance from the equatorial plane
(see caption). 
The first column corresponds to 
$\delta B_{\rm d} = 30\,\mu$G, the second to $\delta B_{\rm d} =
100\,\mu$G, and the third to $\delta B_{\rm d} = 400\,\mu$G. 
In the first row, $z_0=10^{17}\,$cm, and in the second, $z_0=6\times10^{17}\,$cm.
The value of $\eta_{\rm crit}$ in each panel of Fig.~\ref{Spectra_Bands} 
is then: 5.2 (upper left), 17 (upper center), 69 (upper right), 
13 (lower left), 42 (lower center), and 170 (lower right). The downstream 
turbulence level $\delta B_{\rm d}/B_{\rm d}$ can be deduced at any given 
$|z|$ by noting that it is equal to 
$\delta B_{\rm d}/(B_{\rm d,0} \times (|z|/z_{0})) = \eta_{\rm crit}/(|z|/z_{\rm crit})$.
In every panel, all spectra are normalized to the same (arbitrary)
level. The solid red line for the electrons injected at $|z|/z_{0}
\leq 0.02$ dominates over all other lines. A larger fraction of these
particles is accelerated than is the case for injection at larger
$\left|z\right|$, and their spectrum is also harder. 
These results
unambiguously confirm that electron acceleration to high energies
preferentially happens for particles injected at small $|z|$, 
in line with the qualitative discussion in 
Sect.~\ref{Individual_Traj}. 
As is visible in the lower right panel in
Fig.~\ref{Traj_Ele_Pos}, electrons injected at larger $|z|$ move
towards the equatorial plane due to shock-drift, but most of them are
advected into the downstream after a few cycles. Only a small fraction
of them reaches the equatorial region, and this fraction decreases
with the value of $|z|$ at injection. For instance, in the upper left
panel in Fig.~\ref{Spectra_Bands}, the hard high-energy tail of the
dashed orange spectrum for $0.02<|z|/z_{0} \leq 0.04$ 
($3.46 <|z|/z_{\rm crit} \leq 6.93$; $\eta_{\rm crit}=5.2$) is due to those
few particles that have reached the equatorial region and are
subsequently accelerated there. Indeed, this spectrum has about the
same slope as the solid red one. The electrons that do not reach the
equatorial region still gain some energy from their few
shock crossings, and from shock-drift because the average 
change per cycle in $\left|z\right|$ is negative. This is the origin of the small energy
gains experienced by particles injected at higher $|z|$, and of their
\lq\lq bump-like\rq\rq\/ spectra with low-energy cutoffs. See, for
example, the spectra for $0.04<|z|/z_{0} \leq 0.06$ (dotted green
lines) in the first column of Fig.~\ref{Spectra_Bands}, and those for
$0.08<|z|/z_{0} \leq 0.10$ (solid magenta lines) in the second
column. 

For values of $|z|/z_{0}$ larger than those plotted in
Fig.~\ref{Spectra_Bands}, the turbulence level
$\delta B_{\rm d}/B_{\rm d}$ is so low that almost all injected electrons are
advected away into the downstream and do not perform even a single cycle.

Comparing the three columns of Fig.~\ref{Spectra_Bands}, we also
note that the width $z_{\rm w}$ of the favorable region where
electrons can be accelerated to high energies grows with $\delta
B_{\rm d}$. This is unsurprising, because larger turbulence amplitudes
in the downstream correspond to wider regions around the equatorial
plane where the downstream turbulence levels $\delta B_{\rm d}/B_{\rm
  d}$ are sufficiently large for electrons to be scattered back into
the upstream and be accelerated via the first-order Fermi mechanism. For instance,
for $\delta B_{\rm d} = 30\,\mu$G and $z_{0} = 10^{17}$\,cm (upper
left panel), none of the electrons injected in the band
$0.04<|z|/z_{0} \leq 0.06$ is accelerated to high energy. However,
for $\delta B_{\rm d} = 400\,\mu$G (upper right panel), acceleration
in this band is almost as successful as for the central band with
$|z|/z_{0} \leq 0.02$. Indeed, the band 
$0.04<|z|/z_{0} \leq 0.06$ corresponds to a region with downstream 
turbulence levels of $0.5 < \delta B_{\rm d}/B_{\rm d} \leq 0.75$ in the upper
left panel where $\eta_{\rm crit} = 5.2$, and to 
$6.7 < \delta B_{\rm d}/B_{\rm d} \leq 10$ in the upper right
one where $\eta_{\rm crit} = 69$.

By comparing the two rows in Fig.~\ref{Spectra_Bands}, one can see
that the relative width $z_{\rm w}/z_{0}$ of the favorable region for
electron acceleration decreases with $z_{0}$ at $\delta B_{\rm d}$
fixed. For example, for $z_{0} = 10^{17}$\,cm and $\delta B_{\rm d} =
30\,\mu$G (upper left panel), acceleration to high energies still
takes place in the band $0.02<|z|/z_{0} \leq 0.04$ (dashed orange
line), whereas no acceleration to high energies is recorded in the
same band for $z_{0} = 6 \times 10^{17}$\,cm (lower left panel), 
even though the downstream turbulence levels are the same in this band
in both panels. Therefore, the width $z_{\rm w}$ of the 
favorable region for particle acceleration 
does not grow linearly with $z_{0}$. It grows more slowly, 
roughly as $z_{\rm crit}$ and thence as 
$\sqrt{z_{0}}$ (i.e., $z_{\rm w}/z_{0} \propto 1/\sqrt{z_{0}}$).
The band $0.02<|z|/z_{0} \leq 0.04$ corresponds to 
$3.46 <|z|/z_{\rm crit} \leq 6.93$ in the upper left panel, and to 
$8.48 <|z|/z_{\rm crit} \leq 17.0$ in the lower left panel. Indeed, 
the results for this band in the upper panel are similar 
to those in the band $|z|/z_{0} \leq 0.02$ in the lower panel, 
which corresponds to $|z|/z_{\rm crit} \leq 8.48$.

\begin{deluxetable}{ccccc}
\tablecolumns{5}
\tablecaption{Fraction of accelerated electrons over the whole TS \label{Table_Whole_TS}}
\tablehead{\colhead{$z_{0}/(10^{17}\,{\rm cm})$} & \colhead{$\delta B_{\rm d}/(1\,\mu{\rm G})$} & \colhead{$\eta_{\rm crit}$} & \colhead{$\epsilon_{\rm acc,7}$} & \colhead{$\mathcal{F}_{\rm inj}$} \vspace{0.03cm}}
\startdata 
1 & 0.6 & 0.10 & $3.19 \times 10^{-4}$ & 0.05 \\
1 & 1   & 0.17 & $2.06 \times 10^{-3}$ & 0.05 \\
1 & 3   & 0.52 & $6.99 \times 10^{-3}$ & 0.05 \\
1 & 10  & 1.7  & $5.72 \times 10^{-3}$ & 0.05 \\
1 & 30  & 5.2  & $1.01 \times 10^{-2}$ & 0.055 \\
1 & 60  & 10   & $7.22 \times 10^{-3}$ & 0.08 \\
1 & 100 & 17   & $5.78 \times 10^{-3}$ & 0.1 \\
1 & 200 & 35   & $7.59 \times 10^{-3}$ & 0.1 \\
1 & 300 & 52   & $6.42 \times 10^{-3}$ & 0.12 \\
1 & 400 & 69   & $5.93 \times 10^{-3}$ & 0.17 \\
6 & 0.6 & 0.25 & $3.24 \times 10^{-4}$ & 0.05 \\
6 & 1   & 0.42 & $1.39 \times 10^{-3}$ & 0.05 \\
6 & 3   & 1.3  & $3.47 \times 10^{-3}$ & 0.05 \\
6 & 10  & 4.2  & $3.56 \times 10^{-3}$ & 0.05 \\
6 & 30  & 13   & $4.15 \times 10^{-3}$ & 0.055 \\
6 & 60  & 25   & $3.48 \times 10^{-3}$ & 0.08 \\
6 & 100 & 42   & $3.33 \times 10^{-3}$ & 0.1 \\
6 & 200 & 85   & $5.84 \times 10^{-3}$ & 0.1 \\
6 & 300 & $1.3 \times 10^{2}$ & $5.26 \times 10^{-3}$ & 0.12 \\
6 & 400 & $1.7 \times 10^{2}$ & $5.27 \times 10^{-3}$ & 0.17
\tablecomments{Electrons are injected at $|z|/z_{0} \leq \mathcal{F}_{\rm inj}$ with 
energy 1\,TeV. The regular field at $z_0$ is $B_{\rm d,0} = 1$\,mG.
$\epsilon_{\rm acc, 7}$ is the fraction of injected particles accelerated to more than
$7\,$TeV.}
\enddata 
\end{deluxetable}

To expedite the simulations, we choose an upper boundary
on the relative size of
the region where injected electrons can be accelerated to high
energies: $|z|/z_{0} \leq \mathcal{F}_{\rm inj}$. A \lq\lq
generous\rq\rq\/ estimate is provided in
the fifth column of Table~\ref{Table_Whole_TS}, for $\delta B_{\rm
  d}$ within the range $(0.6 - 400)\,\mu$G, and for
$z_{0}=10^{17}$\,cm or $z_{0}=6 \times10^{17}$\,cm. We inject $10^{6}$
electrons at the TS, in the region at $|z|/z_{0} \leq \mathcal{F}_{\rm
  inj}$. In the fourth column of Table~\ref{Table_Whole_TS}, we provide
the fraction $\epsilon_{\rm acc,7}$ of these electrons that are
accelerated to energies $E_{\rm d} \geq 7$\,TeV. We use here the
condition $E_{\rm d} \geq 7$\,TeV because our simulations show that
the overall electron spectrum at the TS is well described by a
power-law above this energy. We find that the spectrum below $\approx
7$\,TeV does not look like a perfect power-law, and displays a small
bump due to the particles injected at large $|z|$. This can be seen
qualitatively by summing up by eye the contributions from all bands in
Fig.~\ref{Spectra_Bands}. These fractions $\epsilon_{\rm acc,7}$
depend on $\mathcal{F}_{\rm inj}$, and multiplying them by
$\mathcal{F}_{\rm inj}/z_{0}$ gives the total acceleration efficiency
for the whole TS in the striped wind region, in planar
geometry. As already expected from Fig.~\ref{Spectra_Bands}, the total
acceleration efficiency tends to grow with $\delta B_{\rm d}$. The
values for $\epsilon_{\rm acc,7}$ are smaller than those for
$\epsilon_{\rm acc,3}$ in Table~\ref{Table_Equator} because of the
higher energy threshold (7\,TeV), and because of the larger size of
the studied region.

Finally, we note that positron acceleration, which is 
inefficient in the equatorial plane, shuts off completely 
at larger $|z|$.

\subsection{Synchrotron X-rays from the Crab Nebula}
\label{Results_Synchrotron}

\begin{figure}
  \includegraphics[width=0.49\textwidth]{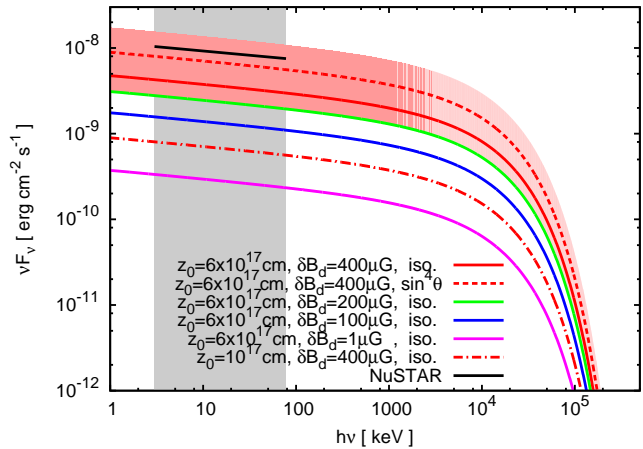}
  \caption{Predicted synchrotron spectra at $h\nu \geq 1$\,keV for the Crab Nebula, versus NuSTAR measurements~\citep[][solid black line]{NuSTAR2015}. Each line corresponds to a different combination of $z_{0}$ and $\delta B_{\rm d}$, for isotropic (\lq\lq iso.\rq\rq\/) or $\propto \sin^{4}\theta$ pulsar winds and for $D_{\rm Crab}=2.0$\,kpc, see the key. The red area is the uncertainty on $\{z_{0}=6\times 10^{17}\,{\rm cm},\,\delta B_{\rm d}=400\,\mu{\rm G}\}$ for $1.5\,{\rm kpc} \leq D_{\rm Crab} \leq 2.5\,{\rm kpc}$.}
\label{Synch_Xrays}
\end{figure}

Using the method described in Sect.~\ref{SynchCalc}, we compute the
synchrotron spectrum, taking $B = 0.5$\,mG for the strength of the
magnetic field in which the electrons cool, and $E_{\max} = 1$\,PeV
for their maximum energy at the TS, cf. Eq.~(\ref{E_max}).  These
values provide a high-energy cutoff in the synchrotron spectrum at
roughly $30$\,MeV, which agrees with observations of the
Crab Nebula, and lies well above the X-ray observations with which we
compare our predictions. The cooling time of electrons of $1$\,PeV is
roughly $10^6$\,s, corresponding to a region of size 
somewhat larger than the acceleration zone considered. The results
of \S~\ref{Whole_TS} show the electron spectrum at the TS to be a
power-law $\propto E^{\alpha_{\rm e}}$ above $E_{\min} = 7$\,TeV
(i.e., $f=7$), which we can expect to extend up to $E_{\max}$. 
The cooling time for electrons of $E_{\rm
  min}$ is roughly $10^8\,$s, corresponding to a size somewhat
smaller than the X-ray nebula, and the energy of the photons emitted
by these electrons is about $1\,$keV, which roughly defines the lower limit of 
the range we attempt to model. 

Observations by NuSTAR~\citep{NuSTAR2015} 
give $\alpha_{\rm e} \simeq -2.2$, which, from
Fig.~\ref{Gamma_e_Equator} (left panel) and Table~\ref{Table_Equator}
implies either $\delta B_{\rm d}> 400\,\mu$G or $>200\,\mu$G for $z_0=
10^{17}\,$cm and $6\times10^{17}\,$cm, respectively. Or,
alternatively, $\delta B_{\rm d}< 1\,\mu$G or $\delta B_{\rm
  d}<0.4\,\mu$G, again for $z_0= 10^{17}\,$cm and
$6\times10^{17}\,$cm, respectively.  (The case of harder spectra is
discussed in Sect.~\ref{Discussion}.)

Assuming the Crab Nebula to be at a distance $D_{\rm Crab}=2.0$\,kpc
from Earth and that the particle flux from the pulsar is
distributed in latitude in proportion to the wind power, with $n=0$ or
$n=4$, 
we plot in Figure~\ref{Synch_Xrays} the synchrotron
spectra $\nu$F$_{\nu}$ at energies $h\nu \geq 1$\,keV, for these
values of $z_{0}$ and $\delta B_{\rm d}$. The normalization is found
using the values of $\epsilon_{\rm acc,7}$ and $\mathcal{F}_{\rm inj}$
from Table~\ref{Table_Whole_TS}, assuming the equatorial radius of
the TS is $r_{\rm TS}=4.3\times 10^{17}$\,cm, the spin-down luminosity
$L_{\rm s.d.} = 5\times 10^{38}$\,erg\,s$^{-1}$, and the mass-loading
parameter $\mu=2\times 10^6$.  In this figure, the solid black line shows the approximate
level of the NuSTAR data~\citep{NuSTAR2015} in the energy band
$3\,{\rm keV} \leq h\nu \leq 78$\,keV (area shaded in grey). Our
prescription of the electron spectrum below
$E_{\rm d}=7$\,TeV, given in Eq.~(\ref{defineq0}),
influences $\nu$F$_{\nu}$ for
$h\nu \lesssim (2-3)$\,keV. 
Since this is not relevant for NuSTAR data, we extrapolate the 
power-law in this figure to below
7\,TeV. 

The four solid lines in Figure~\ref{Synch_Xrays} are calculated for an
isotropic pulsar wind ($n=0$) with $z_{0}= 6 \times 10^{17}$\,cm,
which corresponds to $\Theta \simeq 80^{\circ}$, i.e., an almost
orthogonal rotator.  These four spectra are computed for $\delta
B_{\rm d} = 1\,\mu$G (magenta line), $\delta B_{\rm d} = 100\,\mu$G
(blue), $\delta B_{\rm d} = 200\,\mu$G (green), and $\delta B_{\rm d}
= 400\,\mu$G (red). As expected, the level of the emission increases
with $\delta B_{\rm d}$. The line for $\delta B_{\rm d} = 400\,\mu$G
is still below the NuSTAR data, but is compatible with it if one takes
into account both the uncertainties on the distance to the Crab Nebula
($\pm 0.5$\,kpc) and those on $L_{\rm s.d.}$. In contrast, small
values of $\Theta$ cannot explain the data. For instance, we show, with
the dash-dotted red line, $\nu$F$_{\nu}$ for $z_{0}= 10^{17}$\,cm
(i.e. $\Theta \simeq 13^{\circ}$) and $\delta B_{\rm d} = 400\,\mu$G:
in this case, the predicted level of the emission is an order of
magnitude below the NuSTAR data. Finally, since the pulsar wind may be
anisotropic, we plot, as an example, the case $n=4$
(cf.~Sect.~\ref{SynchCalc}) with $z_{0}= 6 \times 10^{17}$\,cm and
$\delta B_{\rm d} = 400\,\mu$G (dashed red line). The emission is
enhanced by a factor $\approx 2$ with respect to that for an isotropic
wind with the same parameters, which raises it to the level of the
NuSTAR data.  We represent the uncertainty on the two lines with
$\{z_{0},\,\delta B_{\rm d}\}= \{6 \times 10^{17}\,{\rm cm},\,400\,\mu
{\rm G}\}$, caused by the estimate of the distance $D_{\rm Crab}$ to
the Crab Nebula by the area shaded in red.  This shows that the data
are compatible with the above predictions, as well as with those for
an anisotropic pulsar wind with a lower level of turbulence $\delta
B_{\rm d} = 200\,\mu$G.

In our picture, electrons enter the Fermi acceleration process after
thermalization at the TS to an injection energy $E_{\rm inj,d}\sim \mu
m_{\rm e}c^{2}$. The computations reported 
in Fig.~\ref{Synch_Xrays} are performed with $E_{\rm inj,d}=1\,$TeV, which lies 
at the upper end of the permitted range. Repeating these for  
$E_{\rm inj,d}=10$\,GeV leads to a reduction in the X-ray flux by a factor of approximately
25. This is due to reductions of the normalization of the
electron spectrum in the radiating band, and of the size of the region around the 
equatorial plane where electron acceleration occurs, which are only 
partially compensated by the increased particle flux density compared to the 
total power density in the wind. Thus, in our model, such a low injection energy 
is incompatible with the X-ray observations of the Crab Nebula.

\section{Discussion and perspectives}
\label{Discussion}

The synchrotron spectrum of the Crab Nebula follows a power-law, $\nu
F_{\nu} \propto \nu^{-0.1}$, in the X-ray band, according to
observations by NuSTAR~\citep{NuSTAR2015}. This
corresponds to an accelerated electron spectrum at the TS with
$\alpha_{\rm e} \simeq -2.2$, close to the value $-2.23\pm0.01$
predicted for the first-order Fermi mechanism operating at a parallel,
ultra-relativistic shock in the presence of isotropic pitch-angle
diffusion \citep{kirketal00}. However, though ultra-relativistic, the
TS of the wind of the Crab Pulsar is expected to be
perpendicular, rather than parallel, which has led to suggestions that
the Fermi process cannot provide an explanation of the X-ray spectrum
\citep[e.g., ][]{Olmi2016}. The results presented in \S~\ref{Results} 
use an explicit model of the
magnetic field at the TS to demonstrate that
this mechanism is indeed viable. Physically, the reason is that the
drift of particle orbits along the shock surface tends to focus either
electrons or positrons (depending on the pulsar polarity) into the
equatorial current sheet of the nebula. Here, the toroidal magnetic
field is weak, and the level of turbulence suggested by global MHD
simulations is sufficient to provide the scattering needed for the
Fermi process to be effective.

In contrast to the case of a uniform magnetic field, we find that the
spectral index for the more appropriate equatorial current sheet configuration 
depends on the
amplitude of the turbulence. As can be seen in
Fig.~\ref{Gamma_e_Equator} (left panel), both weak and strong
turbulence lead to $\alpha_{\rm e} \simeq -2.2$, but an
intermediate range exists in which a harder spectrum with $\alpha_{\rm e}
\simeq -1.8$ is predicted. In this connection, \lq\lq weak\rq\rq\ and
\lq\lq strong\rq\rq\ refer to the turbulence level at that height in
the sheet where the gyroradius of an injected particle equals its
distance from the equatorial plane. That is, in terms of the parameter defined in
Eq.~(\ref{etacritdef}), $\eta_{\rm crit}\ll1$ and
$\eta_{\rm crit}\gg1$. In the case of the Crab, only \lq\lq
strong\rq\rq\ turbulence amplitudes and a relatively broad current
sheet --- as determined by the angle between the pulsar's magnetic and
rotation axes --- are compatible with the flux level reported by
NuSTAR.  This conclusion rests on the assumption that the angular
dependence of the particle flux carried by the wind is proportional to
that of the total power. At first sight, it might seem that a scenario
in which the particle flux is more strongly concentrated towards the
equatorial plane would lead to an enhanced X-ray flux, and, therefore,
relax the above constraints.  However, an increase in the equatorial
particle flux corresponds to a decrease in the effective value of
$\mu$, and, therefore, of the injection energy. As noted in
\S~\ref{Results}, this reduces the predicted X-ray flux.
These remarks apply to the 
spatially integrated X-ray flux, and assume a level of turbulence that is constant
in time. In principle, the level of turbulence close to the TS can fluctuate on the timescale of 
months. Our computations predict a harder synchrotron
spectrum when $\eta_{\rm crit}\sim 1$~--~$10$. Thus, 
the high spatial resolution observations by the Chandra X-ray Observatory \citep{Mori2004}, that 
reported a photon spectrum corresponding to $\alpha_{\rm e} \simeq - (1.8 - 2.0)$ very close to 
the equator, may have sampled a lower turbulence level in this region of the Nebula. 

In our model, particles are able to return to the shock because they
propagate in a prescribed field of Gaussian turbulence.  
This approach is motivated by MHD simulations of the global
flow pattern, which show turbulence driven roughly on the scale of the
radius of the TS, with an amplitude comparable to the ambient field
strength outside the current sheet. It implicitly assumes that a
turbulent cascade to smaller length scales develops and fills the
downstream region. We tested both
Kolmogorov ($\mathcal{P}(k)\propto k^{-5/3}$) and Bohm ($\propto
k^{-1}$) spectra, and did not find a significant impact on our
results.  This suggests that the choice of spectrum is not important, 
but we note that our limited dynamical range
($L_{\max}/L_{\min} \sim 100$) does not allow us to firmly rule out
any dependence on $\mathcal{P}(k)$ in the case of $L_{\max}/L_{\min}
\gg 100$. 

On the other hand, in the upstream plasma, any
turbulence present must either be imprinted at the launching point of
the wind, or created by reflected particles and/or waves
\citep{lemoinepelletier10,casseetal13}.  Since
the amplitude of the former is difficult to estimate, and the latter
effect is absent in our test-particle simulations, we performed a
series of checks and verified that our results are unaffected by
either the power-spectrum or the amplitude of the upstream turbulence,
provided the latter does not greatly exceed $\sim
0.1\,\mu$G. Complete neglect of the upstream turbulence, on the other
hand, would introduce an unphysical artifact into our simulations,
since a planar 1D treatment without upstream turbulence permits some
particles on Speiser orbits to propagate to arbitrarily large distance
upstream.  In a more realistic picture, such orbits are eliminated by
effects such as irregularities in the incoming wave and radiation
losses of the particles, as well as the spherical geometry appropriate
for a pulsar wind.

The main argument against Fermi acceleration as the mechanism responsible for
producing the X-ray emitting electrons in the Crab Nebula is 
based on the results of PIC simulations \citep{sironispitkovsky09}, 
which show efficient acceleration at
relativistic shocks only when the ambient field is approximately
parallel to the shock normal and the magnetization parameter $\sigma$
is small (typically $<10^{-3}$). 
Because such conditions are expected on only
a very small fraction ($\lesssim 1\%$) of the TS, through which a
correspondingly small fraction of the wind power flows,
particles accelerated there cannot carry the power needed to explain the observed 
X-ray emission \citep{amato14}. 
However,
currently available PIC simulations specify an initially uniform
magnetic field, so that particles can return to the shock only by
scattering on self-generated turbulence. In contrast, the scattering
in our approach results from a turbulent field generated externally by
the global flow pattern.  The region of the TS in which particles are
injected into the acceleration process reaches, in this case, a height
of several times $z_{\rm crit}$ above the equator, corresponding to a
few percent of the area of the TS. The majority ($\gtrsim 90\%$) of
the electrons carried by the wind do not enter the Fermi acceleration
process. Although we do not address the fate of these electrons here,
it is conceivable that another acceleration mechanism operates upon
them, and may be responsible for the radio to optical emission of the
Nebula \citep{Olmi2016}. It is important to note that during
the course of Fermi acceleration, the area of the TS sampled by the
particles grows in proportion to their energy. Therefore, although the
number of participating particles is restricted to those entering
through a few percent of the TS area, the available power is a much
larger fraction of the wind luminosity, and is ultimately sufficient
to produce the observed X-ray flux.

\section{Summary and conclusions}
\label{Conclusions}

Using a global model of the magnetic field, we study the acceleration of 
electrons and positrons at the termination shock of a striped pulsar wind,
and compute the resulting high-energy synchrotron
emission.
For parameters appropriate for the Crab Nebula, we 
find that either electrons or positrons --- but not both --- 
can be accelerated to
$\sim$~PeV energies via the first-order Fermi mechanism in a
ring-shaped region of the TS, around the equatorial plane of the
pulsar. The width of this ring grows with the downstream turbulence
level. The Fermi mechanism shuts off outside
this region because of the strong toroidal field at higher
latitudes. Drifts along the surface of the TS focus the
accelerating particles towards the equatorial plane, and maintain
them on Speiser orbits around it. This favors 
acceleration via the first-order Fermi mechanism, because it 
causes them to cross the TS and re-enter
downstream near this plane, where the toroidal field is weakest and
the turbulence level is largest. In contrast, drifts along the shock 
push particles of the disfavored charge away from this region, thus
hampering their acceleration. The sign of charge that is accelerated
depends on the pulsar polarity. 
Interestingly, modeling of the multi-wavelength emission of the Crab Nebula
suggests that the particles responsible for X-ray emission 
are indeed accelerated close to the equatorial
plane~\citep{Olmi2016}.

The predicted spectral index of the accelerated particles is in the range
$\alpha_{\rm e} \simeq -1.8$ to $-2.4$, and depends on the downstream
turbulence level, being primarily determined by
the electron return probability from the downstream to the upstream,
cf. Fig.~\ref{Gamma_e_Equator}. For turbulence levels $\eta_{\rm crit}
\ll 1$ or $\gg 10$ --- cf.\ Eqs.~(\ref{etacritdef}) and~(\ref{z_crit})
--- we find that $\alpha_{\rm e} \simeq -2.2$, which is consistent with
the photon index $\Gamma = 2.1$ measured for the Crab Nebula in $1 -
100$\,keV X-rays \citep{NuSTAR2015}. The observed X-ray
flux can be reproduced for $\eta_{\rm crit} \gg 10$, provided the
angle between the magnetic and rotation axes of the pulsar is
sufficiently large, cf.~Fig.~\ref{Synch_Xrays}. 
The electron spectrum
hardens to $\alpha_{\rm e} \simeq -1.8$ to $-2.0$ when $\eta_{\rm
  crit} \approx 1 - 10$, which may explain
the hard photon index $\Gamma \simeq 1.9$ to $2.0$ observed by the 
Chandra X-ray Observatory in the central regions of the Crab Nebula~\citep{Mori2004}.
Taking account of the dependence of the spectral index on the level of turbulence
($\eta_{\rm crit}$) may also offer an explanation of the X-ray emission of
other pulsar-wind nebulae.

\acknowledgements

We thank Uri Keshet for useful discussions. This research was
supported by a Grant from the GIF, the German-Israeli Foundation for
Scientific Research and Development.

\appendix

\section{Influence of the grid size on the particle spectra}
\label{appendix}

\begin{figure}
  \centerline{\includegraphics[width=0.49\textwidth]{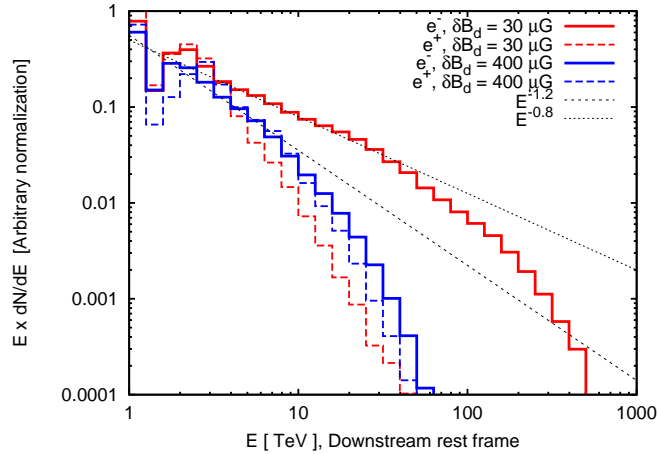}}
  \caption{Simulations using the same parameters as in 
    Fig.~\ref{Spectra_Ele_Pos_Equator} (left panel), but with a
    smaller grid for the turbulent magnetic field: $\mathcal{N}=64$
    and $L_{\max}/L_{\min}=32$, instead of $\mathcal{N}=256$ and
    $L_{\max}/L_{\min}=128$, and a value of $L_{\rm max}$ reduced by a
    factor of four.}
\label{Smaller_Grid}
\end{figure}

We assert in \S~\ref{Results_Equatorial_Plane} that the $\sim (100 - 300)$\,TeV 
cutoffs in the electron spectra of
Figs.~\ref{Spectra_Ele_Pos_Equator} and~\ref{Spectra_Bands} are
artifacts of our simulation technique, caused by the finite
dynamical range $L_{\max}/L_{\min} = 128$ of the turbulence, whereas
the cutoffs that appear at lower energies in the positron spectra of
Fig.~\ref{Spectra_Ele_Pos_Equator} (left panel) and in the spectra
of the electrons injected at large $|z|/z_{0}$ in
Fig.~\ref{Spectra_Bands} are physical. We have confirmed this
interpretation by performing simulations with turbulence generated
on a smaller grid, using a correspondingly reduced value of $L_{\rm max}$. 
For example, Fig.~\ref{Smaller_Grid} shows the electron and
positron spectra for the same parameters as in
Fig.~\ref{Spectra_Ele_Pos_Equator} (left panel), except that the
turbulence is generated on a grid of size $\mathcal{N}=64$ (instead
of $\mathcal{N}=256$), and the value of $L_{\max}$ is reduced by a
factor of four.  By comparing these two figures (which use the same
line types and colors), one sees that, apart from statistical
fluctuations, the positron spectra are identical, whereas the
electron spectra in Fig.~\ref{Smaller_Grid} have a high-energy
cutoff at an energy that is approximately four times smaller than in
Fig.~\ref{Spectra_Ele_Pos_Equator} (left panel).


\bibliographystyle{apj}
\bibliography{references}


\end{document}